\documentclass[a4paper]{article}

\usepackage[a4paper,top=2.5cm,bottom=2.5cm,left=3cm,right=3cm,marginparwidth=1.75cm]{geometry}

\usepackage{amsmath}
\usepackage{amssymb}
\usepackage{authblk}
\usepackage{chngcntr}
\usepackage{csvsimple}
\usepackage{graphicx}
\usepackage{lscape}
\usepackage{subcaption}
\usepackage[section]{placeins}
\usepackage[ruled,vlined]{algorithm2e}
\usepackage[colorinlistoftodos]{todonotes}

\RequirePackage[authoryear,round]{natbib}
\usepackage[colorlinks=true, allcolors=blue]{hyperref}

\newcommand{\norm}[1]{\left\lVert#1\right\rVert}
\newcommand{\iid}{\stackrel{iid}{\sim}}
\newcommand{\ind}{\stackrel{ind}{\sim}}
\newcommand{\tr}{\text{tr} \;}
\newcommand{\var}{\text{Var} \;}
\DeclareMathOperator*{\argmin}{arg\,min}

\title{Active Matrix Factorization for Surveys}
\author[1]{Chelsea Zhang}
\author[2]{Sean J. Taylor}
\author[2]{Curtiss Cobb}
\author[1]{Jasjeet Sekhon}
\affil[1]{UC Berkeley}
\affil[2]{Facebook}

\begin{document}
\maketitle

Amid historically low response rates, survey researchers seek ways to reduce respondent burden while measuring desired concepts with precision. We propose to ask fewer questions of respondents and impute missing responses via probabilistic matrix factorization. A variance-minimizing active learning criterion chooses the most informative questions per respondent. In simulations of our matrix sampling procedure on real-world surveys, as well as a Facebook survey experiment, we find active question selection achieves efficiency gains over baselines. The reduction in imputation error is heterogeneous across questions, and depends on the latent concepts they capture. The imputation procedure can benefit from incorporating respondent side information, modeling responses as ordered logit rather than Gaussian, and accounting for order effects. With our method, survey researchers obtain principled suggestions of questions to retain and, if desired, can automate the design of shorter instruments.

\section{Introduction}

\subsection{Reducing response burden in surveys}

Modern surveys suffer from declining response rates, which inflate administration costs and cast doubt on the validity of inferences. Research has long suspected that survey length plays a role. An inverse association between length and response rate appears in several meta-analyses of mail surveys [\cite{heberlein1978factors, yammarino1991understanding, edwards2002increasing}]. An experiment evaluating redesigns of the U.S. Census found that shortening the questionnaire increased response rate [\cite{dillman1993effects}]. Another experiment showed a sizable negative effect of length on completion in web surveys [\cite{marcus2007compensating}]. There is disagreement about the direction and size of effect [\cite{munger1988use, sheehan2001mail}], although variation in reported effect sizes may be due to disparate survey modes and measures of length [\cite{fan2010factors}]. 

In addition to nonresponse, a longer instrument may be more susceptible to measurement error. Respondents may avoid the cognitive burden of surveys by taking mental shortcuts, such as selecting ``don't know'' or arbitrary responses, a behavior called satisficing [\cite{krosnick1991response}]. Satisficing in the form of straight-line responding -- giving identical answers to consecutive items -- occurs more on a long instrument than a short one [\cite{herzog1981effects}]. There is evidence that both interviewers and interviewees deliberately shortcut interviews to reduce burden, such as answering initial questions in the negative to avoid follow-up questions [\cite{tourangeau2015motivated}]. 

To combat these issues, survey practitioners have suggested reducing respondent burden by asking fewer questions [\cite{kreuter2013facing}]. This idea arose in an earlier era of survey research, when norms shifted from in-person to phone surveys; it became easier for contacts to prematurely end a survey by hanging up. Researchers adapted by administering shorter phone surveys [\cite{groves2011three}]. Assigning a subset of questions to each respondent, otherwise known as matrix sampling, may increase response rate and reduce nonresponse bias [\cite{munger1988use}]. Matrix sampling was implemented for the Consumer Expenditure Interview Survey: respondents were adaptively assigned to sub-questionnaires in a way designed to minimize the variances of mean estimates across expenditure types [\cite{gonzalez2008adaptive}]. In a more fine-grained example of adaptivity, \cite{early2017dynamic} propose to choose questions sequentially to maximize information gain traded off with dropout probability. They also review adaptive design in other fields and simulate their dynamic question-ordering strategies on real-world surveys.

These matrix sampling procedures create missingness in the response matrix consisting of responses to all potential questions. Estimation of marginal population quantities such as question means can proceed using only the responses present, often with weighting adjustment. However, in order to use complete-data methods without discarding data or make downstream decisions for individuals based on their potential responses, missing responses must be imputed. Multiple imputation is a common approach [\cite{rubin2004multiple, thomas2006evaluation, reiter2007multiple}]. It has been argued that surveys should quantify information content using a measure of imputation uncertainty rather than nonresponse rate [\cite{wagner2010fraction}].

We take an imputation approach, leveraging the modern framework of matrix completion. Well studied in the recommender systems literature to predict user-item preferences from sparse ratings across thousands of items, matrix completion enjoys theoretical guarantees and efficient algorithms. Recent work in causal inference uses matrix completion to impute counterfactual outcomes in panel data [\cite{athey2018matrix}]. \cite{kallus2018causal} estimate latent confounders from an incomplete, noisy matrix of covariates using low-rank matrix factorization. Multiple authors have suggested applying matrix completion to survey imputation [\cite{candes2009exact, davenport20141, klopp2015adaptive, josse2016missmda}], but published applications to real-world surveys are rare. One exception is concurrent work by \cite{sengupta2018simple} that examines the predictive ability of matrix completion on survey responses collected by different elicitation strategies.

We envision shortening a burdensome survey by making a wide response matrix sparse. The ability to predict missing from observed responses presupposes a low-dimensional latent structure, which holds in practice as survey items are often correlated. Thus, ours is a natural setting for matrix completion. In addition, the latent quantities from matrix factorization help us prioritize survey items.

\subsection{Optimal design and active learning}

Classical research in several literatures considers how to sample for maximal information gain or minimum-variance parameter estimates. The survey literature has typically focused on optimizing inclusion probabilities of units in sampling frames. Optimal inclusion probabilities have been derived for a variety of survey designs [\cite{sarndal2003model}]. The field of optimal experimental design (OED) considers its decision variables to be design points, from which responses are gathered for parameter estimation. OED minimizes the asymptotic variance of the maximum likelihood estimate, also known as the inverse Fisher information. In general this is a matrix. Different measures of the inverse information matrix define different design criteria: A-optimality minimizes the trace, D-optimality minimizes the determinant and E-optimality minimizes the maximum eigenvalue. 

The definition of decision variables in these optimization problems could expand to survey items or their inclusion probabilities. Indeed, adaptive matrix sampling in \cite{gonzalez2008adaptive} finds assignment probabilities to sub-questionnaires using A- and D-optimality to combine the variances they want to control. Optimal design criteria also feature in the related task of optimal subsampling -- choosing subsets of training data when training a model on the full dataset is too computationally demanding. For instance, optimal subsampling weights have been derived using A-optimality to minimize the variance of the subsample maximum likelihood estimate for logistic regression [\cite{wang2018optimal}].

Optimal design is a principled form of active learning, which encompasses many strategies for training point selection when label acquisition is expensive. \cite{mackay1992information} distinguishes between the goal of OED -- obtaining maximal information about model parameters -- and that of maximizing model performance in a region of input space. For the latter objective, a common baseline is uncertainty sampling, which iteratively chooses the point with greatest predictive uncertainty. Uncertainty sampling is myopic: it does not account for the global effect of item selection on the model. A more principled approach is to choose the point that minimizes the variance component of generalization error -- the predictive variance integrated over the input distribution. \cite{cohn1996active} derive, for several models, closed-form expressions for this integrated variance given a new training point, which can be optimized to suggest the next query point. 

The above active learning strategies, implemented sequentially, produce greedy algorithms. Optimal query points for a multi-step search horizon can be found with a branch-and-bound strategy; theoretical results imply unbounded gains over the greedy strategy, but empirical results show marginal gains [\cite{garnett2012bayesian}].

A theory-to-practice gap may exist for active learning in general. Theoretical results show active learning has lower sample complexity than passive sampling in certain settings, usually within binary classification [\cite{settles2009active}]. In one example, data is distributed uniformly on the unit sphere, and the base learner is a linear separator through the origin [\cite{dasgupta2005analysis}]. However, when the learner is inhomogeneous, the advantage of active learning disappears; it is recovered by weakening the definition of sample complexity [\cite{balcan2010true}]. \cite{attenberg2011inactive} point out several challenges to the adoption of active learning in practice. These include choosing an initial base learner and query selection strategy within the label budget; poor query selection by non-robust strategies, especially with rare classes or concepts; and artificial advantages given to active learning in research experiments.

\subsection{Active learning for matrix factorization}

Active learning strategies have been specialized to matrix factorization to query the most informative entries in the response matrix. The usual setting is a recommender system; the researcher seeks accurate predictions of user ratings of unseen movies. One baseline strategy simply prompts for the most popular items, since users are more likely to recognize them and remain attentive [\cite{elahi2016survey}]. Uncertainty sampling can be used with various models of unobserved matrix entries, such as the graphical lasso and ensembles [\cite{chakraborty2013active}] as well as probabilistic matrix factorization [\cite{sutherland2013active}]. 

Other strategies consider the global effect of item selection. Influence-based strategies find the item that would produce the greatest change in predictions [\cite{rubens2007influence}] or user factors [\cite{karimi2011non}]. Actively querying matrix entries in linearly independent columns, a form of adaptive Nystrom sampling, has been analyzed for completion of symmetric positive semidefinite matrices [\cite{bhargava2017active}]. \cite{silva2012active} maximize mutual information between selected and unobserved instances. For computational efficiency, they also suggest a form of uncertainty sampling in latent space: query users and items with the greatest approximate posterior variance, as measured by the trace.

Still other strategies order items by how much they would reduce the total prediction error of matrix factorization [\cite{golbandi2010bootstrapping, karimi2011towards}]. Direct minimization of RMSE or MAE relies on assumptions about the empirical rating distribution, such as stationarity, since responses are not known before querying. Beyond prediction, active learning for recommender systems could target objectives like profitability or user satisfaction [\cite{rubens2015active, sutherland2013active}].

\subsection{Computerized adaptive testing}

Active learning for matrix factorization could be recast as adaptive item selection that places respondents on latent scales with maximal precision. The literature on item response theory has long pursued this goal. Computerized adaptive testing (CAT) algorithms choose questions online to precisely estimate an individual's latent ability within a fixed question budget. CAT is natural to study in Bayesian terms: responses update the posterior distribution of ability parameters. \cite{montgomery2013computerized} advocate for applying CAT methods to public opinion surveys. They model correctly answering political knowledge questions with logistic regression using a one-dimensional latent ability parameter. Their item selection strategy, which minimizes expected posterior variance of ability, can shorten a battery by 40\% while retaining measurement accuracy.

Multidimensional adaptive testing (MAT) generalizes optimal latent ability estimation to higher dimensions [\cite{segall2009principles}]. The log odds of a correct response is determined by the inner product of multivariate normal ability parameters and fixed ability discrimination parameters. Since the logistic form prevents exact updating of the user ability posterior, a Laplace approximation is used. Segall selects the D-optimal question, which maximizes the determinant of the precision matrix, or equivalently minimizes the size of the posterior credibility region.

In a non-Bayesian approach to MAT, \cite{mulder2009a} arrive at the same matrix following the usual optimal design reasoning: it is the Fisher information of ability parameters. They note that the trace of inverse information includes the determinant as a factor, so A- and D-optimality should act similarly. Their simulations show A- and D-optimality outperform a random selection baseline, while E-optimality is worse than random. In a separate paper adopting the Bayesian approach to MAT, \cite{mulder2009b} analyze additional item selection criteria based on KL divergence and mutual information.

Like \cite{montgomery2013computerized}, we argue that item response theory can inform design of adaptive surveys. In our case, its treatment of low-rank latent structure is particularly relevant. Item selection for MAT is exactly analogous to optimal design for estimating user factors in matrix factorization. 

In the sequel we develop a principled procedure for survey practitioners seeking to shorten a survey. We use active learning to pick the questions to keep and matrix factorization to impute responses for the rest. Modeling responses probabilistically as Gaussian, we obtain an optimally designed offline question order that is interpretable. Modeling responses with the nonconjugate ordered logit likelihood and using approximate inference, we obtain an adaptive question order per respondent. We demonstrate the improved imputation ability of active question selection on left-out questions in multiple survey simulations. Additionally, we confirm this advantage in a Facebook survey experiment comparing active to random and expert-designed question order.

\section{Active matrix completion}

\subsection{Matrix completion methods}

Given $n$ users and $k$ questions, let $R$ denote the $n \times k$ response matrix. Matrix factorization finds a low-rank decomposition of $R$: a set of user factors $U = [u_1, \ldots, u_n]^T$ and question factors $V^T = [v_1, \ldots, v_k]$ such that $R \approx U V^T$. Let $r$ be the dimensionality of latent space, typically small. Then $u_i, v_j \in \mathbb{R}^r$ for all $i$ and $j$.

When $R$ is partially observed, matrix completion adapts matrix factorization to approximately reconstruct observed entries while predicting missing entries. Let $I$ be an indicator matrix for whether the corresponding responses in $R$ exist. $I_{ij} = 1$ implies user $i$ responded to question $j$ with value $R_{ij}$. Matrix completion finds $U$ and $V$ that minimize the reconstruction error $u_i^T v_j$ for $R_{ij}$ on the set $\{(i,j): I_{ij} = 1\}$.

The formulation of matrix completion that enforces a hard rank constraint is nonconvex and generally intractable [\cite{fithian2013flexible}]. It is common to work with a convex relaxation that instead regularizes the nuclear norm, or sum of singular values [\cite{srebro2005maximum}]. This optimization problem seeks a matrix $Z$, in place of $UV^T$, that minimizes reconstruction error; it encourages a low-rank solution by favoring sparsity in the singular values.
\begin{equation}
\min_Z  \frac{1}{2} \sum_{i=1}^n \sum_{j=1}^k I_{ij}(R_{ij}-Z_{ij})^2 + \lambda \norm{Z}_*
\label{eqn:soft-impute}
\end{equation}

This nuclear norm regularized problem enjoys theoretical guarantees: recovery of the complete matrix occurs with high probability when $O(n\, \text{polylog}(n))$ entries are observed at random, with or without noise [\cite{recht2011simpler, negahban2012restricted}]. Moreover, (\ref{eqn:soft-impute}) has an efficient solution in the SoftImpute algorithm by \cite{mazumder2010spectral}. SoftImpute iteratively computes the SVD of $Z$, soft-thresholds the singular values, and updates the entries where $I_{ij} = 0$ with the prediction from the soft-thresholded SVD, until convergence. Hence the solution can be expressed as $Z = UDV^T$ for some matrices $U \in \mathbb{R}^{n \times r}, D \in \mathbb{R}^{r \times r}, V \in \mathbb{R}^{k \times r}$.

An alternate formulation of matrix completion, introduced by \cite{rennie2005fast}, penalizes the Frobenius norm of $U$ and $V$:
\begin{equation} 
\min_{U, V} \frac{1}{2} \sum_{i=1}^n \sum_{j=1}^k I_{ij}(R_{ij}-u_i^T v_j)^2 + \frac{\lambda_U}{2}\norm{U}_F^2 + \frac{\lambda_V}{2}\norm{V}_F^2
\label{eqn:frobenius}
\end{equation}

Problem (\ref{eqn:frobenius}) is nonconvex in $U$ and $V$; it is solved via gradient descent or alternating least squares [\cite{hastie2015matrix}]. This formulation is useful for large-scale problems with low rank, since it is less expensive to operate on $U$ and $V$ than $Z$. The solutions to (\ref{eqn:soft-impute}) and (\ref{eqn:frobenius}) coincide if $\lambda_U = \lambda_V$ and the solution to (\ref{eqn:soft-impute}) has rank at most $r$, due to an identity relating the nuclear norm and sum of Frobenius norms [\cite{fithian2013flexible}].

The user and question factors resulting from either optimization are point estimates, as are the imputed survey responses. We seek a strategy for actively selecting the next survey question based on the uncertainty reduction achieved. To quantify uncertainty over imputed responses, we turn to probabilistic matrix factorization methods.

\subsection{Probabilistic matrix factorization}

Probabilistic matrix factorization (PMF) by \cite{salakhutdinov2008bayesian} models user and question factors as independently normally distributed. Responses add zero-mean, constant-variance Gaussian noise to the inner product of user and question factors. Following the original notation,
\begin{align*}
u_i &\iid \mathcal{N}(\mu_U, \Lambda_U^{-1}) \\
v_j &\iid \mathcal{N}(\mu_V, \Lambda_V^{-1}) \\
R_{ij} \mid U, V &\ind \mathcal{N}(u_i^T v_j, \alpha^{-1})
\end{align*}

With zero-mean, isotropic priors, MAP estimation for PMF corresponds to solving the Frobenius norm regularized problem (\ref{eqn:frobenius}). Specifically, for $\mu_U = \mu_V = 0$, $\Lambda_U = \alpha_U I$ and $\Lambda_V = \alpha_V I$, the MAP estimate of $U$ and $V$ conditional on $R$ is the solution to (\ref{eqn:frobenius}) with regularization parameters $\lambda_U = \alpha_U / \alpha$ and $\lambda_V = \alpha_V / \alpha$.

Bayesian probabilistic matrix factorization (BPMF) places additional normal-Wishart priors on the hyperparameters $\mu_U, \Lambda_U, \mu_V, \Lambda_V$ [\cite{salakhutdinov2008bayesian}]. Posterior inference is performed by Gibbs sampling. They derive the complete conditional for $u_i$ as follows:
\begin{align*}
P(u_i \mid R, V, \mu_U, \Lambda_U, \alpha) &\propto P(u_i, R_{i\cdot} \mid V, \mu_U, \Lambda_U, \alpha) \\
&= \prod_{j=1}^k \left[P(R_{ij} \mid u_i^T v_j, \alpha^{-1})\right]^{I_{ij}} P(u_i \mid \mu_U, \Lambda_U^{-1})
\end{align*}
The complete conditional is conjugate normal with mean $\mu_i^*$ and precision $\Lambda_i^*$:
\begin{align*}
u_i \mid R, V, \mu_U, \Lambda_U, \alpha &\sim \mathcal{N}(\mu_i^*, [\Lambda_i^*]^{-1}) \\
\Lambda_i^* &= \Lambda_U + \alpha \sum_{j=1}^k I_{ij} v_j v_j^T \\
\mu_i^* &= \left[\Lambda_i^*\right]^{-1} \left( \alpha \sum_{j=1}^k I_{ij} R_{ij} v_j + \Lambda_U \mu_U \right)
\end{align*}
This can be recognized as the posterior for Bayesian linear regression with a Gaussian prior and Gaussian noise: $u_i$ are the coefficients, observed rows of $V$ form the design matrix, and $\alpha^{-1}$ is the noise variance. The expression for $\mu_i^*$ also arises in MAP estimation for PMF with zero-mean, isotropic priors; it is the coordinate ascent update for $u_i$ (see section \ref{sec:bpmf-map-relation}). The complete conditional for $v_j$ involves analogous expressions.

\subsection{Active learning formulation}

Assume for simplicity we have a fixed question bank with known factors $v_1, \ldots, v_k$, learned from abundant existing data. A new user $i$ enters the survey pool. We want to select questions optimally for learning $u_i$. For now we do not use side information about users. 

The PMF model admits a convenient online formulation for updating our knowledge about $u_i$ given responses from this user. Suppose, after $t$ responses, $u_i$ is Gaussian with mean $\mu_i^{(t)}$ and variance $\left[\Lambda_i^{(t)}\right]^{-1}$. Next the user answers question $j$. The posterior for $u_i$ is 
\begin{align}
u_i^{(t+1)} \mid R^{(t+1)}, V, \mu_U, \Lambda_U, \alpha &\sim \mathcal{N}\left(\mu_i^{(t+1)}, \left[\Lambda_i^{(t+1)}\right]^{-1}\right) \nonumber \\
\Lambda_i^{(t+1)} &= \Lambda_i^{(t)} + \alpha v_j v_j^T \\
\mu_i^{(t+1)} &= \left[\Lambda_i^{(t+1)}\right]^{-1} \left( \alpha R_{ij} v_j + \Lambda_i^{(t)} \mu_i^{(t)} \right)
\end{align}

We consider how to choose question $j$ optimally. Inspired by approaches in active learning and item response theory, we maximize a measure of posterior information, or minimize a measure of posterior variance. We focus on the trace -- the sum of posterior variance along latent directions. To select the $(t+1)$th question, we solve
\begin{equation}
\min_{j} \tr \left[\Lambda_i^{(t+1)}\right]^{-1}
\label{eqn:active-pmf}
\end{equation}

Our A-optimal criterion is related to minimizing predictive variance. For simplicity, let $\mu(j)$ and $\Sigma(j)$ denote the posterior mean and variance of $u_i$ after asking question $j$. Let $\mathbb{P}$ be the uniform distribution on the unit sphere. Suppose we draw a new question $\tilde v \sim \mathbb{P}$ independently of $u_i$. Our prediction of the response, $u_i^T \tilde v$, has variance $r^{-1} \left[\tr \Sigma(j) + \norm{\mu(j)}_2^2\right]$. See \ref{sec:trace-predvar-relation} for details. Since the second term is nonnegative, minimizing $\tr \Sigma(j)$ corresponds to minimizing a lower bound on the predictive variance along uniform latent directions.

For more intuition, we rewrite our optimization problem, letting $\lambda_\ell(\cdot)$ denote the $\ell$th eigenvalue of a matrix. (\ref{eqn:active-pmf}) is equivalent to
\[ \min_j \sum_{\ell=1}^r \lambda_\ell\left(\left[\Lambda_i^{(t+1)}\right]^{-1}\right) = \min_j \sum_{\ell=1}^r \left[\lambda_\ell \left(\Lambda_i^{(t+1)}\right) \right]^{-1} \]

This variance criterion penalizes small eigenvalues of the precision matrix, corresponding to directions in latent space with least information. Information is acquired by sampling questions with factors $v_j$ that lie in those directions. The optimal sampling strategy chooses questions as a function of their informativeness and their contribution to less explored directions. For a spherical prior, provided questions exist in many directions with similar magnitudes, the strategy prefers new questions roughly orthogonal to previous questions.

Algorithm \ref{algo:active} summarizes our active strategy. Note some limitations of this simple version. First, the algorithm is greedy, selecting only one question at a time. We could extend the search horizon with dynamic programming or strategies that do not exhaustively enumerate the search space. Second, the optimal sequence of questions can be computed offline, as the objective in (\ref{eqn:active-pmf}) does not depend on response values. There is one active question order for all respondents. This unrealistic property results from the assumptions of Gaussianity and fixed $V$, which produced the closed-form user posterior update.

\begin{algorithm}
\DontPrintSemicolon
\SetKwInOut{Given}{given}
\Given{question factors $v_1, \ldots, v_k$\\ 
prior parameters $\mu_U, \Lambda_U$\\ 
noise variance $\alpha^{-1}$\\
desired survey length $T$}
\BlankLine
Set unasked questions: $\, \mathcal{U} \leftarrow \{1, \ldots, k\}$\;
Set user prior: $\mu_0 \leftarrow \mu_U,\; \Lambda_0 \leftarrow \Lambda_U$\;
\For{$t\leftarrow 1$ \KwTo $T$}{
  Choose next question, $j \leftarrow \argmin_{\ell \in \mathcal{U}} \tr \left[ \left( \Lambda_{t-1} + \alpha v_\ell v_\ell^T \right)^{-1} \right]$\;
  Collect response $R_{ij}$\;
  \BlankLine
  Update user posterior:\;
  $\Lambda_t \leftarrow \Lambda_{t-1} + \alpha v_j v_j^T$\;
  $\mu_t \leftarrow \Lambda_t^{-1} \left( \alpha R_{ij} v_j + \Lambda_{t-1} \mu_{t-1} \right)$\;
  \BlankLine
  Mark question asked: $\mathcal{U} \leftarrow \mathcal{U} \setminus \{j\}$\;
}
\caption{Active question selection for single user} \label{algo:active}
\end{algorithm}

\section{Data and evaluation methods}

\subsection{Datasets}

\begin{table}
\centering
\begin{tabular}{|l|c|c|c|}
\hline
Dataset & Number of respondents & Number of questions \\
\hline
Facebook on-platform survey & 11793 & 53 \\
CCES 2012 & 54535 & 29 \\
CCES 2016 & 64600 & 38 \\
CCES 2016 (full) & 64600 & 61 \\
CCES 2018 & 60000 & 42 \\
\hline
\end{tabular}
\caption{Dataset characteristics. CCES 2016 (full) refers to CCES 2016 with extra covariates. \label{tbl:datasets}}
\end{table}

We simulate active question selection on multiple datasets, summarized in Table \ref{tbl:datasets}. The Facebook survey is a survey of Facebook users, administered on the app or web interface, with a variety of questions about their experiences with the product and the company. The Facebook on-platform survey was administered in random order.

The Cooperative Congressional Election Survey (CCES) is a national Internet survey of adult U.S. citizens conducted by YouGov that seeks to gauge voter opinions about prevailing political issues and elected officials, before and after an election [\cite{ansolabehere2010cces}]. Respondents are selected by matching an opt-in respondent pool to a stratified random sample from the American Community Survey. Our main results use the pre-election surveys from 2016, limiting consideration to Common Content questions that ask respondents to evaluate national political issues or entities on a binary or ordinal scale. For robustness checks we expand the question set to include voter demographics, party identification and other characteristics; we refer to this as the ``full'' CCES dataset. We exclude questions about voter actions in the past year and opinions of state or local representatives, as well as questions with a majority of responses missing.

\begin{figure}
  \begin{subfigure}[b]{0.49\textwidth}
    \includegraphics[width=\textwidth]{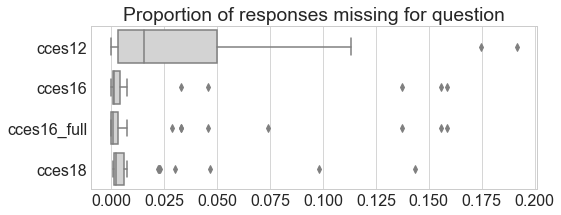}
  \end{subfigure}
  \begin{subfigure}[b]{0.49\textwidth}
    \includegraphics[width=\textwidth]{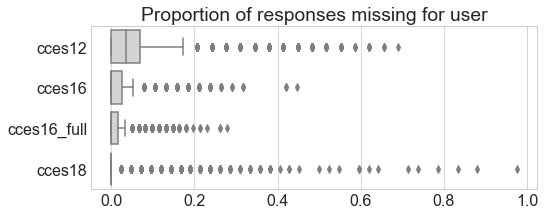}
  \end{subfigure}
  \caption{Missingness distribution of datasets at the question and user levels \label{fig:dataset-sparsity}}
\end{figure}

For each survey question, allowable responses are rescaled to $[-1,1]$. Some responses will be missing, either because they were not present in the original dataset, or because we dropped response values that violated the ordinal assumption. CCES has low overall missingness rates: 3.7\% in 2012, 1.5\% in 2016 and 1.2\% in 2018. The missingness distributions by question and by user are shown in Figure \ref{fig:dataset-sparsity}.

\subsection{Simulating the active strategy} \label{sec:sim-procedure}

Our simulations begin by randomly splitting the respondent set into a training half and a simulation half. We perform matrix factorization on the training responses to estimate $V$. We choose SoftImpute for this step due to its efficiency and empirical stability of $V$ across simulations. We use the SoftImpute implementation in the \texttt{fancyimpute} package [\cite{rubinsteyn2016fancyimpute}]. The regularization parameter $\lambda$ is selected by grid search with warm starts as recommended by \cite{mazumder2010spectral}, based on mean absolute error on a 20\% validation set within the training half.

On the simulation half, we hold out a set of responses and simulate running the survey on the remaining responses. We select the next question per respondent using the active strategy, reveal available responses to that question, and update each user posterior. We predict held-out responses using the estimated $V$ and the MAP estimate for all user factors. We repeat this process until all questions have been asked. Since the active strategy is greedy, we can truncate the process at any point to obtain the actively chosen questions for a given survey length.

All simulations compare the active strategy to a baseline of asking questions in a random order per respondent and, in the case of CCES, existing question order. We focus on mean absolute error (MAE) and bias of predictions. We also compute mean squared error and the proportion of predictions with the wrong sign. 

We determine the holdout set in two ways. The first method reserves a random 20\% of each user's responses, effectively punching holes in the response matrix. We call this the ``sparse'' holdout set. It allows us to evaluate error averaged over questions and make summary comparisons of question selection strategies. The sparse holdout set has drawbacks: the artifice that the simulation procedure treats these responses as missing when the active or random strategy requests them; and higher variability in per-question evaluation error due to using one-fifth of responses. Thus, our second holdout method is leave-one-question-out (LOOCV) cross-validation, used to evaluate prediction error for individual questions. For each question, we simulate the survey on the response matrix with that column removed. LOOCV uses all available responses for a question to evaluate its imputation error, but repeats the survey once per question. As LOOCV is more computationally demanding, we rely on the sparse holdout set to evaluate variations on the simulation procedure quickly.

One variation is to estimate question factors $v_j$ by solving the Frobenius norm regularized problem (\ref{eqn:frobenius}) rather than SoftImpute, due to the connection between (\ref{eqn:frobenius}) and MAP estimation for PMF with simple priors. However, this nonconvex optimization yields highly variable question factors and orderings. With SoftImpute, question factors are relatively stable across simulations, up to sign changes. Another variation is to minimize measures of posterior variance other than the trace, like the determinant and maximum eigenvalue. These optimal design criteria have similar overall predictive performance and active orderings.

Our main results use a rank-4 matrix decomposition ($r=4$). SoftImpute solves the nuclear norm regularized problem subject to this hard rank constraint. For any value of $r$, we search for $\lambda$ as above. Setting $r=4$ results in lower prediction error than $r=2$, while keeping the dimensionality of latent space manageable. A higher-rank decomposition ($r=8$) does not reduce prediction error further. Greater $r$ requires greater $\lambda$ to avoid overfitting; this may shrink the highest-variance components more than necessary.

In the active strategy, we set the prior mean $\mu_U$ and prior precision $\Lambda_U$ using empirical Bayes. Specifically, we set $\mu_U$ and $\Lambda_U^{-1}$ to the sample mean and covariance of the rows of $UD$, the implied user factors from SoftImpute. It remains to set the noise variance $\alpha^{-1}$. By Popoviciu's inequality and the prior rescaling of responses to $[-1,1]$, we know $\alpha^{-1} \leq 1$. Our main results use the upper bound $(\alpha^{-1} = 1)$, though we tried smaller values. Future work should estimate $\alpha$ from the training half of responses.

Results from these alternate simulation settings appear in Appendix \ref{sec:tune-sim-params}.

\section{Results}
\subsection{Does the active strategy impute more efficiently?}

Below we showcase the imputation ability of simulated survey strategies on the 2016 CCES. Results for other years and the Facebook survey appear in Appendix \ref{sec:cces18-results}, \ref{sec:cces12-results} and \ref{sec:facebook-results}. Replication code is available \href{https://github.com/chelseaz/active_survey}{here}.

\begin{figure}[tb]
  \centering
  \begin{subfigure}[b]{0.55\textwidth}
    \includegraphics[width=\textwidth]{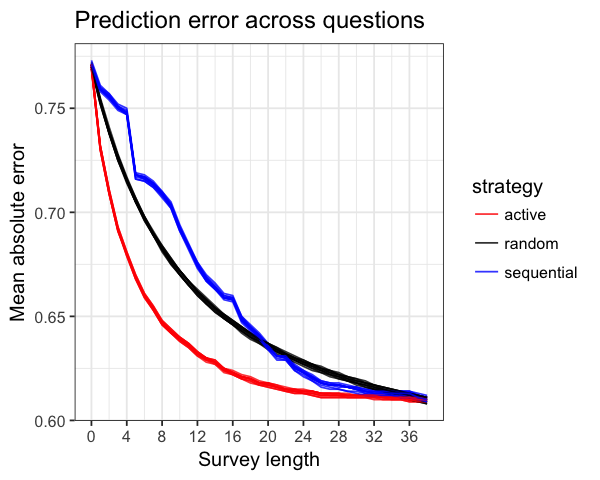}
    \caption{Prediction error on the sparse holdout set, measured across 10 iterations of simulating the 2016 CCES. We show MAE averaged over all questions.}
    \label{fig:cces16-compare-metrics}
  \end{subfigure}
  \hfill
  \begin{subfigure}[b]{0.40\textwidth}
    \includegraphics[width=\textwidth]{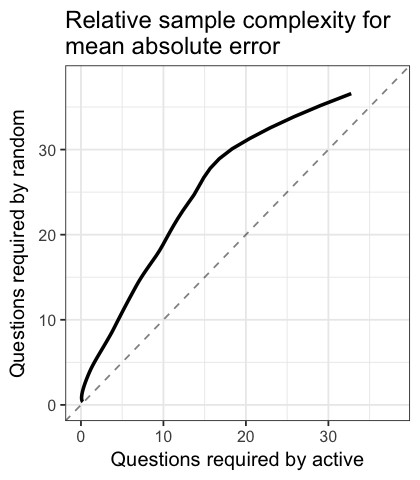}
    \caption{Sample complexity of the active strategy relative to random-order questions for the 2016 CCES.}
    \label{fig:cces16-relative-complexity}
  \end{subfigure}
  \caption{We present summary measures of imputation error for simulated surveys using each question selection strategy. The source of randomness is the sparse holdout set, conditional on the training/simulation split. Each iteration of each strategy uses a different sparse holdout set. Figure \ref{fig:cces16-relative-complexity} is derived from Figure \ref{fig:cces16-compare-metrics} as follows: the solid line plots the number of questions required by each strategy to attain the same level of error. For instance, the active strategy requires 5 questions to reach the same MAE as the random strategy with 10 questions. When the curve lies above the dashed $45^{\circ}$ line, active sampling outperforms random. The curve is obtained by fitting, for strategy $s$, a loess smoother $f_s(\epsilon)$ to predict number of questions required for error level $\epsilon$. We plot $(f_{random}(\epsilon), f_{active}(\epsilon))$ for the range of $\epsilon$ attained by both strategies.}
\end{figure}

We first examine overall performance on the sparse holdout set over multiple simulations. Predictions with actively chosen questions outperform predictions with randomly or sequentially chosen questions (Figure \ref{fig:cces16-compare-metrics}). The active strategy attains lower imputation error, averaged over questions, for simulated surveys of short or medium length. Only when two-thirds of the questions have been asked do the strategies converge in overall performance; this error level is the minimum achievable by low-rank matrix factorization on this dataset. Similar dynamics for other error measures appear in Figure \ref{fig:cces16-compare-metrics-extra}.

Another measure of efficiency gain is sample complexity -- the number of responses required by each strategy to reach a given error level. The active strategy almost always requires fewer responses (Figure \ref{fig:cces16-relative-complexity}). Suppose we ask 20 questions in a random order for each respondent; the active strategy reaches the same imputation quality with nearly half as many questions.

The active strategy minimizes the optimal design criterion, as Figure \ref{fig:cces16-objective} verifies. It is advisable to balance exploration and exploitation when optimizing under uncertainty. We introduce exploration with a simple $\epsilon$-greedy modification to the active strategy: choose a random question with probability $\epsilon = 0.05$; otherwise choose the A-optimal question. $\epsilon$-greedy question selection does not outperform active question selection in terms of prediction error (Figure \ref{fig:cces16-epsilon-greedy-compare-metrics}). Since our simulations treat question factors as fixed, exploration cannot reduce their estimation error. Henceforth we focus on the active ($\epsilon=0$) strategy, with the caveat that some form of exploration is preferable when question factors contain nontrivial uncertainty.

\begin{figure}[tb]
  \centering
  \begin{subfigure}{0.48\textwidth}
    \includegraphics[width=\textwidth]{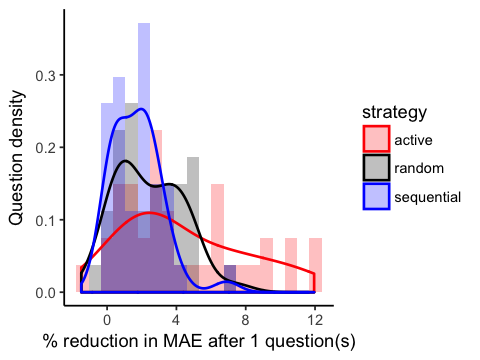}
  \end{subfigure}
  \hfill
  \begin{subfigure}{0.48\textwidth}
    \includegraphics[width=\textwidth]{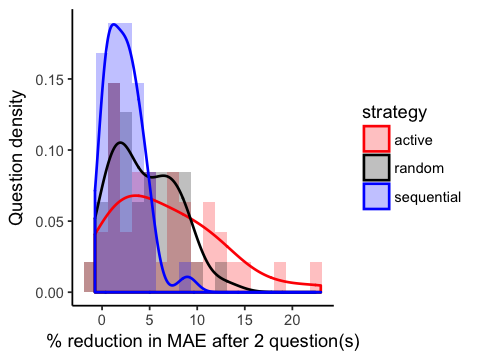}
  \end{subfigure}
  \vskip 5mm
  \begin{subfigure}{0.48\textwidth}
    \includegraphics[width=\textwidth]{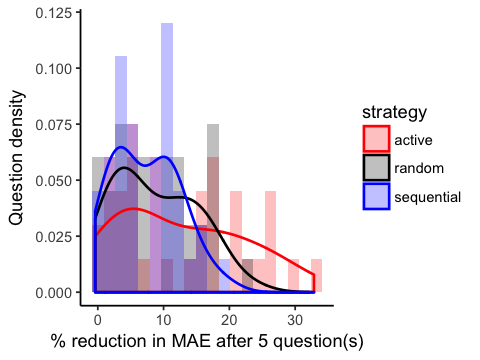}
  \end{subfigure}
  \hfill
  \begin{subfigure}{0.48\textwidth}
    \includegraphics[width=\textwidth]{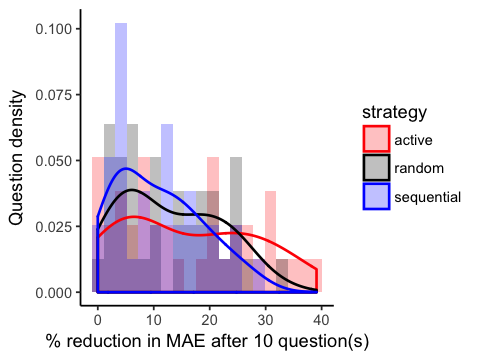}
  \end{subfigure}
  \caption{Reduction in imputation error from pre-survey levels for the 2016 CCES. We show percent reduction in mean absolute error as a distribution over questions, smoothed by kernel density estimation. Error per question is evaluated by leave-one-out cross-validation.}
  \label{fig:cces16-error-reduction}
\end{figure}

Evaluation metrics on the sparse holdout set mask considerable heterogeneity in which questions are amenable to imputation and which benefit from active selection. Figure \ref{fig:cces16-error-reduction} plots the reduction in LOOCV error per question. More questions experience at least a 20\% reduction in imputation error after a five-question active survey, compared to five randomly or originally ordered questions. The advantage of the active strategy is apparent after one or two questions. As survey length increases, all strategies achieve greater error reduction, and this advantage narrows. The error reduction of a single actively chosen question is more notable for the 2018 CCES: at least 12\% on half of that year's question set (Figure \ref{fig:cces18-error-reduction}). Political preferences, at least those captured by the CCES, have become more predictable if we know what to ask.


\begin{figure}
\centering
\includegraphics[width=\textwidth]{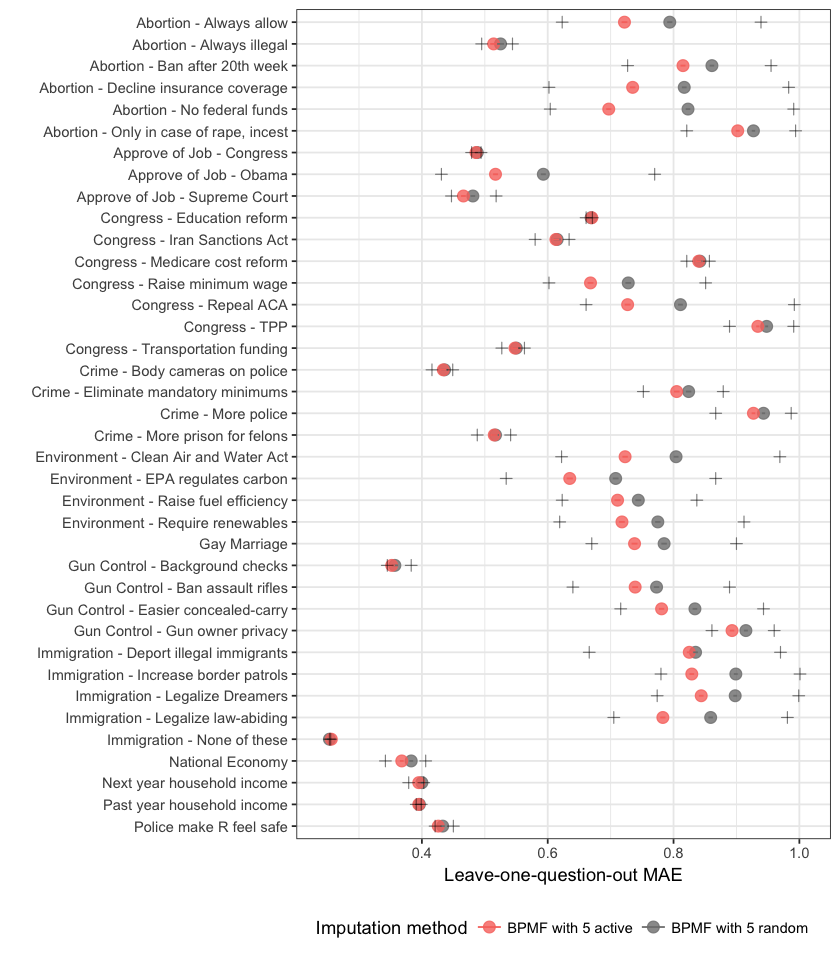}
\caption{Mean absolute prediction error per question after a short survey using active or random question selection. We simulate the survey with one question held out, reveal responses to five actively or randomly chosen questions, and predict responses to the held-out question. We repeat this for each question to produce leave-one-out cross-validation error. $2\sigma$ confidence intervals are not visible due to the number of respondents in the simulation half. The gray ``+'' bounds represent prediction error under the pre-survey condition of no knowledge (right) and the oracle condition of knowing all other responses (left).}
\label{fig:cces16-compare-questions}
\end{figure}

Figure \ref{fig:cces16-compare-questions} locates the questions for which the imputation abilities of active and random selection diverge. As the next section shows, the questions that the active strategy helps are correlated with the actively chosen questions. The bounds in Figure \ref{fig:cces16-compare-questions} indicate what extent of error reduction is possible from knowing no responses (``pre-survey'') to knowing all available responses (``oracle''). The oracle bound quantifies the irreducible error of imputing each question with low-rank matrix factorization. Some questions are inherently harder to impute -- their oracle MAE is close to 1. Other questions have low oracle error and low pre-survey error. In both cases, whether question selection is active or random makes little difference. For intuition, on a binary question with possible responses $\{-1,1\}$, MAE of 1 is achievable by (i) randomly guessing -1 or 1 with equal probability or (ii) always predicting 0. 

One component of irreducible error is bias. The pre-survey and oracle bias per question are shown in Figure \ref{fig:cces16-compare-questions-bias}. Revealing all non-held-out responses changes bias little from pre-survey levels; bias is mostly determined once question factors have been estimated. Low-rank matrix factorization produces small bias relative to MAE: less than 0.1 in either direction for all questions, and less than 0.05 for all but two questions. There is no clear relationship between questions with high irreducible error and those with relatively high bias.

The relative irreducible errors are affected by our decision to scale all responses to $[-1, 1]$. Questions with evenly distributed binary responses will tend to have higher irreducible error than those with lopsided binary responses or ordinal responses. The alternative of zero-centered, unit-variance scaling would complicate interpretation, as the response values would depend on the response distribution. Our choice of scaling balances competing goals of interpretability and standardization. 

With this caveat in mind, it may be ineffective to impute questions with high irreducible error. The active strategy has little room to help. The survey researcher is advised to include such questions in the eventual survey, regardless of their position in the active question order, if accurate measurement of these constructs is a priority.

\subsection{Which questions does the active strategy prefer?}

The active strategy ranks questions by precision gained in the latent representation of a user. For more intuition, see Appendix \ref{sec:cces16-2d-viz}, which visualizes the latent representations of users and questions in two dimensions. 

\begin{figure}
\centering
\includegraphics[width=\textwidth]{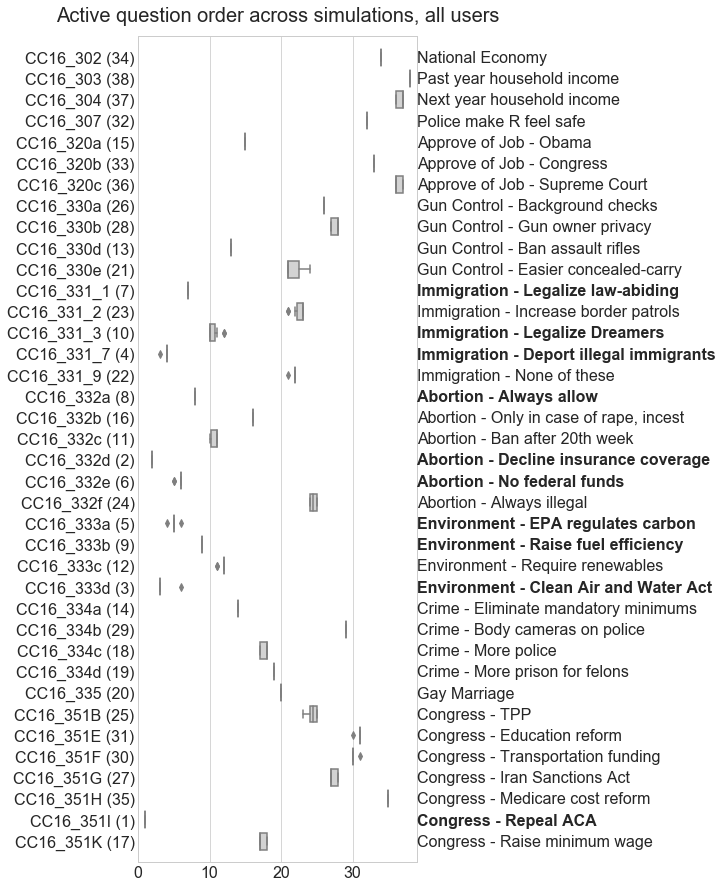}
\caption{Active ordering for 2016 CCES using a rank-4 matrix decomposition with no questions held out. The box plot shows the rank of each question across 10 training/simulation splits. Position in the active ordering, according to median rank, is in parentheses. The top 10 questions appear in bold. \label{fig:cces16-question-rank}}
\end{figure}

Active item selection produces a stable question order (Figure \ref{fig:cces16-question-rank}). The foremost question is whether to repeal the Affordable Care Act (ACA). Questions about immigration, abortion and environmental policies are prioritized: a 10-question active survey includes multiple questions from each topic. These topics receive the most predictive improvement from active selection in a five-question survey (Figure \ref{fig:cces16-compare-questions}). The predictive advantage of active learning comes not from covering these topics exhaustively but from sampling informative items within correlated sets. Active selection also benefits predictions of Obama approval and support for gun restrictions, which do not appear in the top 10. Crime and economic questions appear later in the active ordering; the predictive advantage of active learning on these topics is minimal.

The active ordering for the 2018 and 2012 questionnaires appear in Figures \ref{fig:cces18-question-rank} and \ref{fig:cces12-question-rank}. In 2012, the active strategy favors the ACA and immigration; it passes over abortion- and environment-related questions. These latter issues may feature less in latent concepts due to a shortage of relevant questions that year. The top 10 active questions for 2018 address a hodgepodge of issues, including the top issues from 2016 as well as taxes and trade. The leading question in 2018, whose response yields clear predictive gains, is whether to appoint Brett Kavanaugh to the Supreme Court.

We check the robustness of our 2016 results to our question inclusion criteria for the CCES. We progressively add questions about respondent political affiliation, demographics, education and other characteristics. Questions with categorical responses, like race, are converted into indicators. Note this one-hot encoding artificially creates a separate survey question per response value; multinomial logit modeling would be more appropriate in practice. Appendix \ref{sec:question-order-robustness} contains active orderings with these additional questions.

The active ordering with augmented questions remains largely faithful to the active ordering in Figure \ref{fig:cces16-question-rank}. Though questions about gender, party identification, parenthood and home ownership slot into the first 20 positions, questions about the environment, abortion, and the ACA remain prominent. Gender and Obama approval displace immigration questions from the top 10. Interestingly, the active strategy postpones questions about race and education, possibly because these one-hot-encoded variables are not well captured by a low-rank matrix decomposition.

We interpret the latent concepts for which the active strategy gathers information. Figure \ref{fig:cces16-pcs} displays question factors as loadings on these latent concepts. The first direction broadly indicates partisanship: Democratic and Republican policies tend to have loadings with opposite sign. It makes sense that this principal component contains most prior variance. Partisanship is highly correlated with opinions about the environment, abortion and immigration, which explains their prominence in the active ordering and their improved imputation under the active strategy. That the highest-variance component captures partisanship is replicated in 2018 and 2012 (Figures \ref{fig:cces18-pcs} and \ref{fig:cces12-pcs}, respectively). 

In all three years as well, the second component seems to represent level of bipartisan support. For instance, increased prison sentences for repeat felons, requiring police to wear body cameras and background checks for all gun purchases are broadly popular policies supported by 84\%, 87\% and 90\% of 2016 respondents, respectively. These load highly in the negative direction. Opposite these is a ``none-of-the-above'' question about immigration policies, which only 5\% of respondents supported. In 2018, background checks, North Korea sanctions and provisions of the tax bill command broad support and load opposite the unpopular policy of criminalizing abortion in all circumstances. In 2012, the then-popular Keystone XL pipeline loads opposite the Ryan budget bill and approval of Congress, which was at historic lows.

In 2016, the third principal component correlates support for greener environmental policies, support for abortion restrictions and opposition to gay marriage. This suggests a group of socially conservative or religious respondents who are concerned about the environment. The fourth component correlates support for greener environmental policies and opposition to abortion restrictions with support for tougher crime and immigration policies. The active strategy doubles down on environmental, abortion and immigration questions in order to ascertain membership in these groups. Going beyond two latent dimensions helps to identify parts of the electorate that do not behave according to conventional partisan wisdom.

In other years, the active strategy rounds out a 10-question survey with questions that load primarily on non-partisanship components. These include the Simpson-Bowles budget plan and penalizing employers of illegal immigrants in 2012. In 2012 the third component aligns with support for isolationist and xenophobic policies, while the fourth component prioritizes fiscal issues, namely free trade and deficit reduction. Latent concepts change from year to year. The active strategy adapts and actively seeks information along these time-varying directions.

\section{Side information}

When we incorporate side information about respondents, both active and random strategies may see efficiency gains. Theoretical results have established that sufficiently informative side information improves the sample complexity of matrix completion [\cite{xu2013speedup, chiang2015matrix}]. We give a simple proof of concept of the value of side information for the Facebook survey. We subgroup respondents based on two covariates: country and length of time since joining Facebook. Each simulation user's prior parameters are set to the subgroup mean and covariance in the training half. The resulting active order is still deterministic but specialized to the subgroup. We do not have enough power to determine whether using side information in this way reduces overall prediction error (Figure \ref{fig:facebook-subgroups-compare-metrics}). Future work could expand the covariate set or impose shrinkage across subgroups via hierarchical modeling.

Another way to incorporate side information is to include respondent covariates directly as responses in matrix factorization. Revisiting the full CCES 2016 dataset, we reveal responses to all covariate questions before simulating any survey questions, so that PMF updates each user's prior with these ``free'' covariates. In practice, this can be done by collecting covariates from the sampling frame or at the start of the survey. Simulations show free covariates reduce imputation error early in the survey at the cost of introducing bias (Figure \ref{fig:cces16-free-covariates-compare-metrics}). The information advantage of free covariates disappears as more questions are asked; the active strategy breaks even around 17 questions. Strategies using free covariates have higher oracle error than their agnostic counterparts. The active ordering with free covariates is similar to that without, though covariates again substitute for questions about immigration (Figure \ref{fig:cces16-question-rank-free-covariates}).

\section{Ordered logit response model}

In this section we explore adjusting the model to better capture binary and ordinal response values. We replace the Gaussian likelihood for responses with the ordered logit likelihood. The ordered logit model, also known as the proportional odds model, is prevalent in social science research [\cite{fullerton2012proportional}]. As we will see, this change breaks the determinism of the active order; question selection will now depend on the respondent's previous answers.

Prior work models quantized outputs in the response matrix with a variety of link functions, including logistic, probit and multinomial logit [\cite{davenport20141, cao2015categorical, klopp2015adaptive}]. These works formulate matrix completion as maximum likelihood with nuclear norm regularization. We continue with a probabilistic matrix factorization approach. Posterior inference in the ordered logit model involves non-conjugacy, so we resort to variational inference, also used for matrix completion in [\cite{lim2007variational, seeger2012fast}]. We forfeit the closed-form posterior update exploited by the active strategy for PMF, but the Laplace approximation offers a way forward.

\subsection{Probabilistic matrix factorization model}

Our ordered logit response model retains normal priors for user and question factors. Responses are integer-valued starting at 1. We allow heterogeneity across questions: the number of response values can differ across questions, as can response frequencies. The model becomes:
\begin{align*}
u_i &\iid \mathcal{N}(\mu_U, \Lambda_U^{-1}) \\
v_j &\iid \mathcal{N}(\mu_V, \Lambda_V^{-1}) \\
R_{ij} \mid U, V &\ind \text{OrderedLogit}(u_i^T v_j, \beta_j)
\end{align*}

$R_{ij}$ takes values in the range $\{1, 2, \ldots, M_j\}$, where $M_j$ is the question-specific maximum response value. Let $\pi_{j,m}$ denote the probability that $R_{ij} = m$. The probabilities $\{\pi_{j,1}, \pi_{j,2}, \ldots, \pi_{j,M_j}\}$ are defined by the logistic link and a series of question-specific cutpoints $\beta_j = (\beta_{j,1}, \ldots, \beta_{j,M_j-1})$. For simplicity of presentation, we drop the indexing for question $j$. Thus $R_{ij}$ takes values in $\{1, 2, \ldots, M\}$ with probabilities $\{\pi_1, \pi_2, \ldots, \pi_M\}$, parameterized by cutpoints $\beta = (\beta_1, \ldots, \beta_{M-1})$ as follows:
\begin{align*}
\text{logit}\left(\sum_{k=1}^m \pi_k \right) &= u_i^T v_j + \beta_m \qquad (m = 1, \ldots, M-1) \\
\pi_{M} &= 1-\sum_{k=1}^{M-1} \pi_k
\end{align*}

\subsection{Inference}

We perform posterior inference on $U, V$ in the above model when estimating user and question factors from the training half and after collecting an additional response from each user in the simulation half. Using the updated posteriors per iteration, we compute prediction error on held-out survey responses. See Algorithm \ref{algo:ordlogit}. 

We obtain approximate posteriors for $U$ and $V$ given $R$ in the above model using mean-field variational inference. We implement this in \texttt{edward} [\cite{tran2016edward}]. Our variational distributions are fully factorized Gaussian:
\begin{align*}
q(U) &= \prod_{i=1}^n q(u_i) = \prod_{i=1}^n \prod_{j=1}^r \mathcal{N}(\mu_{ij}, \sigma_{ij}) \\
q(V) &= \prod_{j=1}^k q(v_j) = \prod_{j=1}^k \prod_{i=1}^r \mathcal{N}(\nu_{ji}, \tau_{ji})
\end{align*}

Variational inference finds parameters $\{\mu_{ij}, \sigma_{ij}, \nu_{ji}, \tau_{ji}\}_{i,j}$ that maximize the evidence lower bound, or equivalently minimize the KL divergence between the variational distribution and the true posterior. We employ priors $\mu_U = \mu_V \equiv 0$ and $\Lambda_V = \Lambda_V \equiv \mathbf{I}_r$. To obtain cutpoints $\beta$, we follow the inverse approach in the \texttt{rstanarm} package [\cite{gabry2016rstanarm}]. For each question, we draw from the simplex probabilities $\pi = (\pi_1, \ldots, \pi_M)$ corresponding to the ordinal response values. Specifically, we draw $\pi \sim \text{Dirichlet}(c_1, \ldots, c_M)$, where the concentration parameters are prior counts of the response values. That is, we set $c_m$ equal to the number of times a respondent answers $m$ to this question in the training half. We then apply the logit transform to all but the last entry of $\texttt{cumsum}(\pi)$, obtaining 
\[ \beta_m = \text{logit}\left( \sum_{k=1}^m \pi_k \right) \qquad (m = 1, \ldots, M-1) \]

To predict $R_{ij}$, we set $u_i$ and $v_j$ equal to their variational means $\mu_i$ and $\nu_j$ and compute the mean of the resulting ordered logit random variable.

\subsection{Active learning formulation}

We also update our item selection strategy for the ordered logit response model. In this situation, we consider $V$ to be fixed; we approximate it with the variational means $\{\nu_{j}\}_{j=1}^k$. Consider administering the survey to user $i$. We seek the question $j$ that maximizes information about $u_i$. 

We simplify the ordered logit model to a single user:
\begin{align*}
u_i &\iid \mathcal{N}(\mu_U, \Lambda_U^{-1}) \\
R_{ij} \mid U, V &\ind \text{OrderedLogit}(u_i^T v_j, \beta)
\end{align*}

Unlike in the case of Gaussian likelihood, we do not have a conjugate, closed-form update for the posterior of $u_i$, so we cannot minimize a measure of posterior variance directly. Instead, we work with the variance of the Laplace approximation, or the portion of this variance we can control through item selection -- the Fisher information. This approach follows the optimal design literature, notably \cite{segall2009principles}, who applies it to logistic likelihood for binary responses. Our approach can be considered an ordered logit generalization of \cite{segall2009principles}.

Fisher information is computed around a value of $u_i$. We estimate $u_i$ with the most recent mean of the user variational distribution, $\mu_i$, from probabilistic matrix factorization. Repurposing this provisional estimate of $u_i$ is more computationally efficient than the alternative of computing a MAP estimate of $u_i$ in the single-user model.

We denote the Fisher information gained from a response to question $j$ as $\mathcal{I}^j(u_i)$, and the \textit{observed} Fisher information from observing response $m$ to question $j$ as $\mathcal{J}^j(u_i; m)$. Then
\[ \mathcal{J}^j(u_i; m) = -\frac{\partial^2}{\partial u_i \partial u_i^T} \log \text{Pr}(R_{ij} = m \mid u_i, v_j, \beta) \]
and, letting $\pi_{ijm}$ denote $\text{Pr}(R_{ij} = m \mid u_i, v_j, \beta)$,
\[ \mathcal{I}^j(u_i) = E\left[ \mathcal{J}^j(u_i; R_{ij}) \right] = \sum_{m=1}^M \pi_{ijm} \mathcal{J}^j(u_i; m) \]

In the ordered logit model, $\mathcal{J}^j(u_i; m)$ and $\mathcal{I}^j(u_i)$ involve complicated but closed-form expressions. The Hessians are computed with autodifferentiation in \texttt{edward}.

Let $\mathcal{O}$ contain the indices of past questions and $\mathcal{U}$ the indices of unasked questions. We compute the sum of observed information over $\mathcal{O}$, and consider adding a Fisher information term for question $j \in \mathcal{U}$. We determine which question would contribute the most information to $u_i$ in expectation. More formally, we find the question that minimizes the variance of the Laplace approximation A-optimally:
\begin{equation}
\min_{j \in \mathcal{U}} \tr \left[ \Lambda_U + \sum_{\ell \in \mathcal{O}} \mathcal{J}^\ell(u_i; R_{i\ell}) + \mathcal{I}^j(u_i) \right]^{-1}
\label{eqn:active-ordlogit}
\end{equation}

When expanding the survey by one question, the per-user item selection problems can be solved in parallel. This subprocedure is placed in context in Algorithm \ref{algo:ordlogit}.

\begin{algorithm}
\DontPrintSemicolon
\SetKwFunction{OrdLogitPMF}{ordered\_logit\_matrix\_factorization}
\SetKwInOut{Given}{given}
\Given{training responses $R^{train}$\\
cutpoints $\beta$\\
desired survey length $T$}
\BlankLine
$\{\mu_{ij}, \sigma_{ij}, \nu_{ji}, \tau_{ji}\}_{i,j} \leftarrow \OrdLogitPMF(R^{train}, \beta)$\;
Set $(v_1, \ldots, v_k)$ to variational means $(\nu_1, \ldots, \nu_k)$\;
Set $\Lambda_U^{-1}$ to empirical covariance of $\{\mu_1, \ldots, \mu_n\}$\;
Initialize record of questions per user: $\mathcal{O}_i \leftarrow \emptyset,\; \mathcal{U}_i \leftarrow \{1, \ldots, k\} \; \forall i$\;
Initialize revealed responses $R^{asked}$ to empty $n \times k$ matrix\;
\For{$t\leftarrow 1$ \KwTo $T$}{
  Ask one question of all users:\;
  \For{$i\leftarrow 1$ \KwTo $n$}{
    Choose next question, $j \leftarrow \argmin_{j \in \mathcal{U}} \tr \left[ \Lambda_U + \sum_{\ell \in \mathcal{O}} \mathcal{J}^\ell(u_i; R_{i\ell}) + \mathcal{I}^j(u_i) \right]^{-1}$\;
    Collect response: $R_{ij}^{asked} \leftarrow R_{ij}$\;
    Mark question asked: $\mathcal{O}_i \leftarrow \mathcal{O}_i \cup \{j\}, \; \mathcal{U}_i \leftarrow \mathcal{U}_i \setminus \{j\}$\;
  }
  \BlankLine
  $\{\mu_{ij}, \sigma_{ij}, \nu_{ji}, \tau_{ji}\}_{i,j} \leftarrow \OrdLogitPMF(R^{asked} \cup R^{train}, \beta)$\;
  Set $(u_1, \ldots, u_n)$ to variational means $(\mu_1, \ldots, \mu_n)$\;
  Set $R_{ij}^{pred}$ to mean of $\text{OrderedLogit}(\mu_i^T \nu_j, \beta)$ random variable $\forall i, j $\;
  Compute prediction error of $R^{pred}$ for $R$
}
\caption{Active strategy simulation for ordered logit model} \label{algo:ordlogit}
\end{algorithm}

\subsection{Results}

Our evaluation procedure departs from previous simulations in two ways. First, predictions under the ordered logit model are on the same scale as the original ordinal responses; questions with more allowable responses will tend to have higher error. Thus, we rescale prediction error per question to be comparable to that of PMF, which predicts responses rescaled to $[-1,1]$. Second, inference and item selection are more computationally intensive with the ordered logit model. To avoid the additional computational burden of leave-one-question-out cross-validation, we resort to 5-fold cross-validation on questions. Results with the ordered logit model appear in Appendix \ref{sec:ordered-logit-results}.

We find imputation error under active question selection is reduced faster with ordered logit response modeling than with PMF. This is especially apparent after two questions. The difference after one question comes from the matrix factorization step rather than item selection, since the first actively chosen question is the same under both models. We have modeled responses more appropriately as ordinal and introduced additional parameters in the form of question-specific cutpoints. Item selection with the ordered logit likelihood may contribute to imputation gains starting with the second question; this warrants further investigation.

With five actively selected questions, the ordered logit predictions are nearly at oracle level. Not much room for improvement remains; it may be worth terminating the survey here. The pre-survey and oracle error bounds are close to those of PMF, giving us confidence in the rescaling step. Note that oracle error may exceed mid-survey error in some cases due to the possibility of increased bias with more responses. Indeed, greater bias is a shortcoming of the ordered logit procedure: unlike with PMF, the oracle bias can be considerably higher than the pre-survey bias. This is unsurprising given our reliance on approximate inference. 

With the ordered logit model, there is more variability in the active question order within one simulation than across PMF simulations. We can visualize the active ordering more granularly as individual paths through survey questions (Figure \ref{fig:cces16-ordlogit-question-paths}). The variability in question rank arises not from a few common question orderings, but rather from diverse, personalized paths dependent on responses to previous questions.

Across these individual paths, immigration, abortion and environmental policies are prioritized, and ACA repeal remains the top question across sampled users. These similarities to the PMF active ordering reinforce our earlier findings on question importance. Some questions that appear late in the PMF ordering have highly variable position in the adaptive ordering. These include perception of the U.S. economy over the past year and whether abortion should always be illegal. Approval of Obama is asked anywhere from second to 20th. This was the least stable question in our PMF robustness checks -- it leaps into second place with the addition of covariate questions -- so its variable position under the ordered logit model is natural.

\section{Order effects}

Algorithmic ordering of survey questions may produce biased responses (compared to random ordering) when responses vary according to their position in the survey instrument -- a phenomenon known as \emph{order effects}. In this section we estimate the magnitude of order effects in the randomly-ordered Facebook survey in order to understand how large the bias introduced by the active strategy is likely to be.

\begin{figure}
  \centering
  \begin{subfigure}{0.45\textwidth}
    \includegraphics[width=\textwidth]{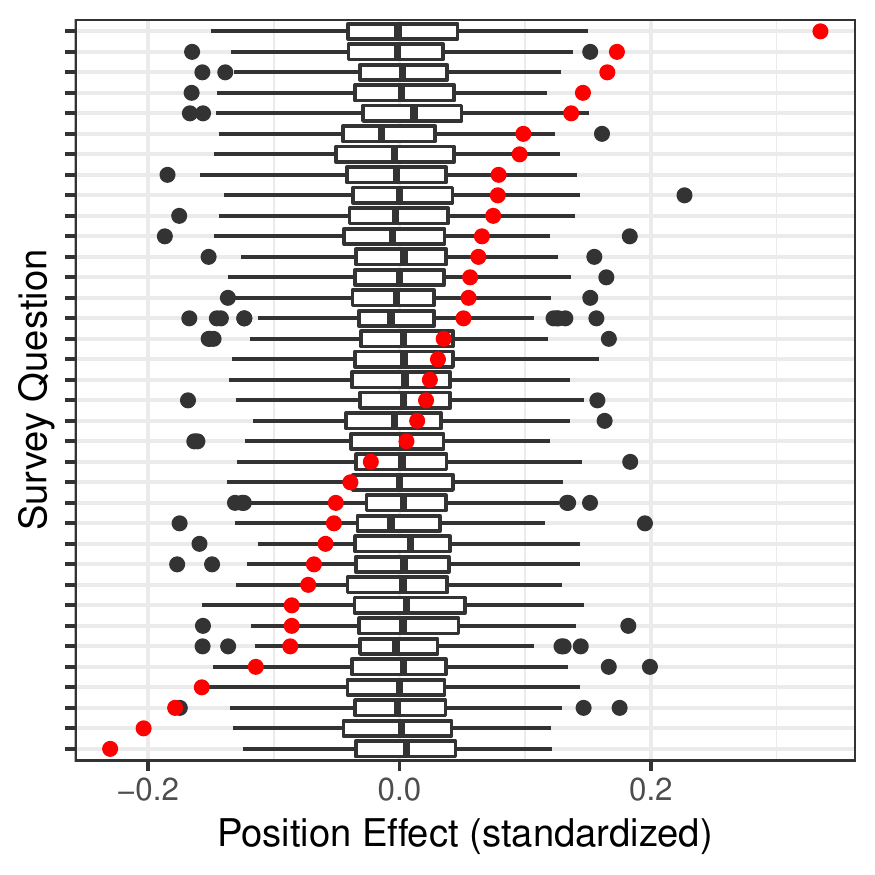}
    \caption{Empirical (red) versus null distribution (boxplots) of question position effects, obtained by permuting question order and fitting linear regressions of standardized survey responses.}
  \end{subfigure}
  \hfill
  \begin{subfigure}{0.50\textwidth}
    \includegraphics[width=\textwidth]{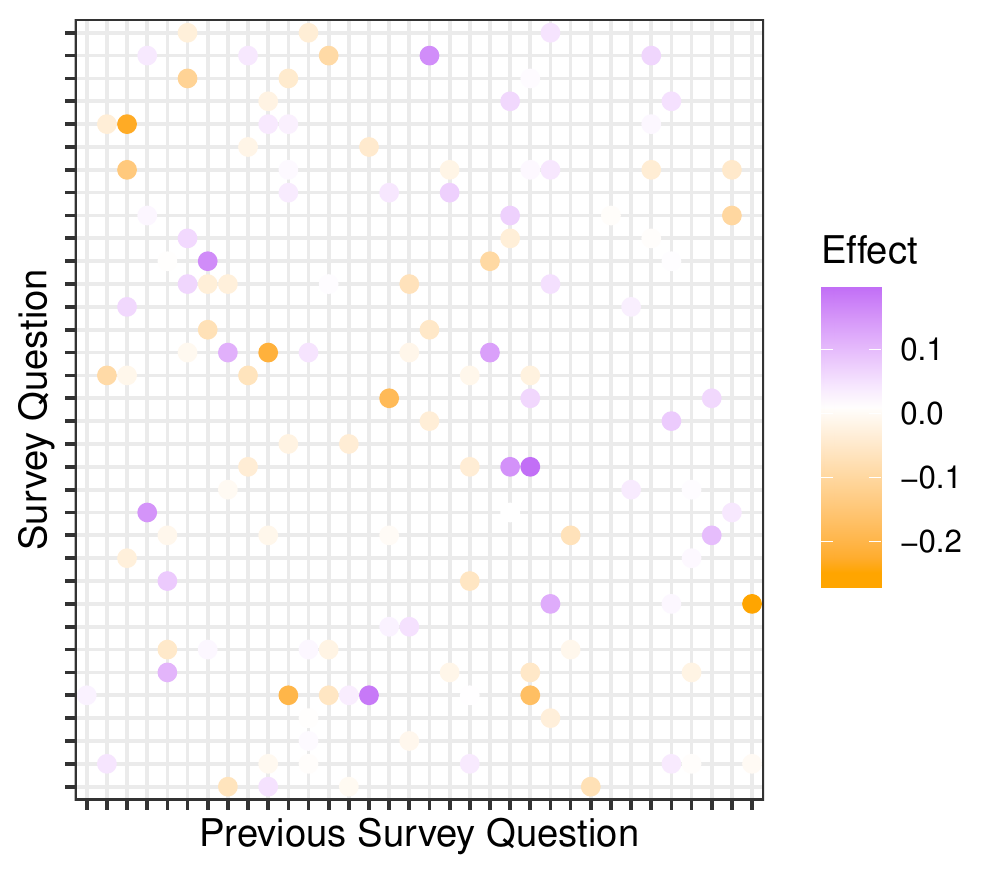}
    \caption{Effect of previous question on next question response for the Facebook survey. To detect large order effects, if any, we estimate these via L1-penalized regression on all pairwise question interactions.}
  \end{subfigure}
  \caption{Estimated order effects in the Facebook survey. \label{fig:question-position-effects}}
\end{figure}

In order to estimate position effects, we fit a linear regression per survey question with the relative position of the question in the order as a predictor.  In this model we use data from only completed surveys (about 30\% of the surveys) in order to preclude attrition bias.  Using this model, we estimate the difference in standardized response for each question appearing at the end of the survey compared to the beginning.  As a null distribution, we randomly re-order the survey and fit the same model $200$ times.  The results are presented in Figure~\ref{fig:question-position-effects}(a).  We
find evidence for a number of survey questions exhibiting position effects, where responses vary significantly depending on whether they are asked early or late in the survey.  The worst-case bias appears to be about $0.3$ standard deviations on the response scale.

We also estimate whether the previous survey question a user answered affects the response on the following question.  We fit an L1-penalized regression with a parameter for all pairs of survey questions and previous questions, using 10-fold cross-validation to select the optimal penalty parameter.  We visualize the results of this model in Figure~\ref{fig:question-position-effects}(b).  About 10\% of the possible question pairs exhibit a non-zero interaction effect. Some questions tend to be influential on the following question (columns with multiple points) while others tend to be more likely to be affected by the prior question (rows with multiple points).  Similarly to position effects, the effects we observe are usually less than $0.2$ standard deviations on the response scale. Moreover, these effects are not robust to the choice of whether to include the pairs beginning at odd- or even-numbered question positions. In sum, we have not detected large, persistent interaction effects. 

While these estimates are rudimentary, they provide some sense of the size of the bias introduced by the active learning algorithm -- it should be a small contribution to the total error compared to the variance reduction we achieve through receiving more informative responses.  In the following section we directly estimate this bias by running the active and random orderings online. Our experiment confirms the order effects for the Facebook survey are small. Note larger order effects could be mitigated by $\epsilon$-greedy question selection or by domain-expert rearrangement of the questions included in an active survey.

\section{Experimental comparison}

Thus far the imputation gains of active question selection have only been shown in simulations that reorder existing survey responses. To verify these gains experimentally, we conducted a second version of the Facebook survey using the deterministic active ordering from rank-4 PMF. Active order was compared to random order and an order generated by a survey methodologist, henceforth called expert order. The expert order was deterministic aside from certain random-order subsets of questions. To reduce burden, we trimmed the survey from 53 to 33 questions by removing some question groupings. Respondents were recruited from four countries and randomly assigned to conditions. The active, random and expert conditions had 4224, 4211 and 4177 respondents, respectively.

We find both active and expert orderings induce order effects relative to random order (Figure \ref{fig:facebook-exp-bias-response}). Where it exists, the bias in mean response is small -- no more than about 0.3 standard deviations on the response scale, consistent with our analysis of the first Facebook survey.

Our primary concern is the imputation performance of short surveys designed by each strategy. As before, our evaluation truncates the survey to a given length and computes the error of PMF predictions on remaining responses under each strategy. In addition, we reuse the notion of pre-survey imputation error, the error of predictions using the user prior. Pre-survey error on a given question can be slightly different across conditions, due to order effects and nonresponse bias (Figure \ref{fig:facebook-exp-bias-pre-error}). We control for these small biases by computing each strategy's reduction from its own pre-survey error. 

The following analysis differs from our simulation-based comparisons in two ways. First, while our simulations held out each question in turn to compute LOOCV error, we respect the order in which experimental responses are gathered. Hence, imputation error after a five-question active survey would be computed for all questions appearing after the first five in the active ordering. For the random and expert orderings, we compute per-question imputation error restricted to respondents who answered this question after the first five. Second, our PMF predictions use question factors estimated from the first Facebook survey. As the surveys occurred a year apart, the latent structure may have changed. Imputation for the survey experiment may improve with updated question factors, and the active order may no longer be optimal.

\begin{figure}
  \centering
  \begin{subfigure}{0.48\textwidth}
    \includegraphics[width=\textwidth]{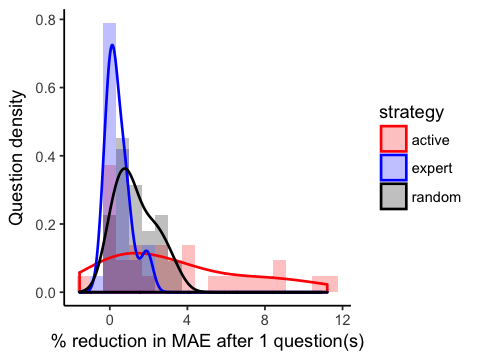}
  \end{subfigure}
  \hfill
  \begin{subfigure}{0.48\textwidth}
    \includegraphics[width=\textwidth]{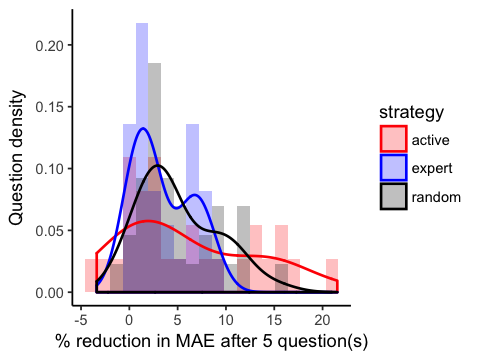}
  \end{subfigure}
  \caption{Reduction in imputation error from pre-survey levels for the Facebook survey experiment. We show percent reduction in mean absolute error as a distribution over questions, smoothed by kernel density estimation. We omit questions with undefined imputation error after the truncated survey, namely the first question in the active order (left) and the first five questions in the active order (right).}
  \label{fig:facebook-exp-error-reduction}
\end{figure}

Still, it is apparent after one question that the active order produces greater error reduction than random and expert order on a subset of questions (Figure \ref{fig:facebook-exp-error-reduction}). Error reduction grows as more questions are asked; at five questions the random and expert order begin to catch up. The empirical distribution of error reduction under the active strategy contains more right tail mass despite omitting the first five actively selected questions, two of which had high error reduction after a one-question survey. For all strategies the distribution of error reduction appears bimodal, suggesting one set of questions is easier to impute than the rest. Error reduction in the experimental data is consistent with reduction in LOOCV error on simulated data (Figure \ref{fig:facebook-exp-versus-simulation}). The questions we expect to benefit actually benefit, and they benefit more from active order in short surveys.

\subsection{Impact of active ordering on nonresponse}

However, the active ordering pays for its imputation advantage with higher nonresponse rates. The active order experiences greater attrition on the initial question and maintains a lower proportion of respondents over the course of the survey (Figure \ref{fig:facebook-exp-respondents-remaining}). A possible explanation is that the active order places more controversial questions upfront. One proxy for controversiality is how often a question is skipped; three of the first four questions in the active order are among the five most skipped by active-order respondents. They are also the three most skipped by expert-order respondents, who see these questions later. 

The expert order experiences the least attrition over the first half of the survey, compensating in part for its slower reduction of imputation error. These findings imply a tradeoff between information gain and dropoff. A future iteration of the survey could represent this tradeoff in the active learning objective. In addition, a wide pilot survey could be undertaken to discover informative questions with low rates of nonresponse.

\section{Discussion and future work}

If we commit to shortening a survey, we will impute uncollected responses with error. We identify questions that drive down imputation error via active learning that maximizes user precision in the latent space of concepts. As we explore the CCES questions preferred by the active strategy, we develop insights about the most informative set of questions for predicting political opinion. The active ordering offers a new notion of feature importance for domains with wide item sets and low-dimensional latent structure.

We have presented two variants of the active strategy. The first produces a deterministic question order, which follows from Gaussian response modeling in PMF. This is convenient for survey applications that require a predictable design upfront. It is also computationally simple. However, the property that future questions do not depend on past responses is unrealistic, as evidenced by the improved imputation ability of the ordered logit response model. With the ordered logit likelihood, the active ordering adapts to collected responses at the price of increased computation.

Both forms of active question selection assume away uncertainty in estimated question factors. Future work should incorporate this uncertainty into the active learning objective, for instance with bandit algorithms that use an upper confidence bound for the optimality criterion or sample from its distribution. The active strategy could also account for temporal uncertainty. It is natural to consider latent concepts as time-varying; explicitly modeling these dynamics could produce an active ordering that evolves automatically.

Future work should devote special attention to the logistics of survey administration. The difficulty of implementing an active ordering depends on the degree of adaptivity. The deterministic order from PMF is relatively straightforward to apply across survey modes; the adaptive order powered by the ordered logit model would require more infrastructure. Web and computer-assisted telephone surveys could lean on backend software to suggest the next question. In-person field surveys would need a mechanism for inputting responses and quickly receiving the next question, such as a mobile app that calls a low-latency API for the active ordering. It would be productive to integrate with existing survey platforms to make the active strategy available. Additionally, suggesting questions in batch may be more practical than sequentially; this calls for optimal design with a multi-step horizon. It is also an opportunity to design logical question groupings informed by domain knowledge of practitioners.

Our evaluation methods focus on the ability to predict individual responses to individual questions. This is useful for a range of applications. For instance, using the CCES, researchers have investigated whether alignment between the policy preferences of constituents and the votes of their legislators predicts constituent support for their legislators [\cite{ansolabehere2013cooperative}]. This analysis requires individual responses to ``roll call'' questions about policy preferences -- the same questions we have included from the CCES. As another example, in survey experiments, responses from a baseline survey could serve as pre-treatment covariates for estimation of heterogeneous treatment effects [\cite{broockman2017design}]. A priori, the analyst cares equally about the quality of imputed responses across respondents and questions. Still, survey researchers are often interested in quantities describing the marginal and joint distributions of specific questions. Future work should evaluate imputation quality and develop active learning in the context of these estimators.

We have utilized side information in two simple ways, but many avenues exist for more sophisticated modeling of side information. One potential approach is to create personalized user priors with multilevel regression. Alternatively, one could map covariates to a user prior with a Gaussian process [\cite{adams2010incorporating, zhou2012kernelized}]. A different tack is to add terms for user and question covariates to the response specification; they enter naturally into the nuclear norm minimization in \cite{athey2018matrix} and the BPMF model in \cite{porteous2010bayesian}. Our simulations with free covariates imply an exchange rate between side information and survey responses. Trading off the information gain and acquisition cost of both in a user-specific way is a design opportunity.

Matrix completion as an imputation procedure can suffer from both nonresponse and model misspecification bias. The standard matrix completion loss assumes entries are missing completely at random. The tenuous plausibility of this assumption is exacerbated by the active strategy, which tailors questions to a user's inferred latent position. Weighting approaches attempt to de-bias the loss by regularizing a weighted nuclear norm [\cite{srebro2010collaborative}] or applying inverse propensity weights to reconstruction terms [\cite{schnabel2016recommendations, athey2018matrix}]. Alternatively, one could model the missing-data mechanism explicitly [\cite{marlin2009collaborative}]. Even with these corrections, a low-rank linear decomposition cannot capture all of the response variance; the remainder is reflected in oracle error. It could be worth exploring nonlinear matrix factorization in the form of Gaussian process latent variable models [\cite{lawrence2009non}].

Of course, it may make most sense to reserve part of the question budget for key items whose direct measurement error is sufficiently lower than the imputation error by matrix factorization. Ultimately, the degree to which active question selection influences survey design rests with the survey researcher. Our method can guide the researcher in creating shorter instruments by suggesting informative questions in a principled manner. At the other end of the spectrum, it can automate adaptive survey design.

\section{Acknowledgements}

We thank JD Astudillo and Felicitas Mittereder for running the Facebook surveys and designing the expert order. We also thank Karan Aggarwal, Eytan Bakshy, George Berry, Dennis Feehan, Avi Feller, Will Fithian, Mike Jordan, Frauke Kreuter, Luke Miratrix, Jacob Montgomery, Adam Obeng, Rebecca Powell and Alex Theodoridis for their valuable input. CZ is supported by an NSF Graduate Research Fellowship. CZ and JS are supported by Office of Naval Research (ONR) Grant N00014-15-1-2367.


\bibliography{main}

\appendix
\counterwithin{figure}{section}
\counterwithin{table}{section}







\section{Proofs}

\subsection{Relationship between PMF posterior mean and MAP estimation} \label{sec:bpmf-map-relation}

The complete conditional for user factors is
\begin{align*}
u_i \mid R, V, \mu_U, \Lambda_U, \alpha &\sim \mathcal{N}(\mu_i^*, [\Lambda_i^*]^{-1}) \\
\Lambda_i^* &= \Lambda_U + \alpha \sum_{j=1}^k I_{ij} v_j v_j^T \\
\mu_i^* &= \left[\Lambda_i^*\right]^{-1} \left( \alpha \sum_{j=1}^k I_{ij} R_{ij} v_j + \Lambda_U \mu_U \right)
\end{align*}
We show this expression for $\mu_i^*$ also arises in MAP estimation for PMF, as the coordinate ascent update for $u_i$.

We reproduce the MAP objective function, Eq (4) in [\cite{salakhutdinov2008bayesian}]. We use prior hyperparameters $\mu_U = \mu_V = 0$, $\Lambda_U = \alpha_U I$, $\Lambda_V = \alpha_V I$ and introduce $\lambda_U = \alpha_U / \alpha$, $\lambda_V = \alpha_V / \alpha$. For these settings, the objective function is simply the Frobenius norm regularized problem.
\[ \ell(U, V, R) = \frac{1}{2} \sum_{i=1}^n \sum_{j=1}^k I_{ij}(R_{ij}-u_i^T v_j)^2 + \frac{\lambda_U}{2} \sum_{i=1}^n \norm{u_i}_2^2 + \frac{\lambda_V}{2} \sum_{j=1}^k \norm{v_j}_2^2 \]
Taking the derivative with respect to $u_i$ and setting to 0 yields
\begin{align*}
\frac{\partial}{\partial u_i} \ell(U, V, R) &= -\sum_{j=1}^k I_{ij}(R_{ij}-u_i^T v_j)v_j^T + \lambda_U u_i^T = 0 \\
\sum_{j=1}^k I_{ij}R_{ij}v_j &= \left(\sum_{j=1}^k I_{ij} v_j v_j^T + \lambda_U I\right) u_i
\end{align*}
The coordinate ascent update is
\begin{align*}
u_i^* &= \left(\sum_{j=1}^k I_{ij} v_j v_j^T + \lambda_U I\right)^{-1} \left(\sum_{j=1}^k I_{ij}R_{ij}v_j\right) \\
&= \left(\alpha \sum_{j=1}^k I_{ij} v_j v_j^T + \alpha_U I\right)^{-1} \left(\alpha \sum_{j=1}^k I_{ij}R_{ij}v_j\right) \\
&= \left[\Lambda_i^*\right]^{-1} \left( \alpha \sum_{j=1}^k I_{ij} R_{ij} v_j + \Lambda_U \mu_U \right) \\
&= \mu_i^*
\end{align*}

\subsection{Relationship between trace minimization and predictive variance} \label{sec:trace-predvar-relation}

Let $\tilde v$ be drawn from $\{v: \norm{v}_2 = 1\}$, the uniform distribution on the unit sphere, independently of $u_i$. We show the predictive variance is
\[ \var(u_i^T \tilde v) = \frac{1}{r} E \norm{u_i}_2^2 = \frac{1}{r}\left(\tr \Sigma + \norm{\mu}_2^2\right) \]
Proof:
\begin{align*}
\var(u_i^T \tilde v) &= E(u_i^T \tilde v \tilde v^T u_i) - \left(E(u_i^T \tilde v)\right)^2 \\
&= E\left(u_i^T E(\tilde v \tilde v^T \mid u_i) u_i\right) - \left(E(u_i)^T E(\tilde v)\right)^2 \\
&= E\left(u_i^T \frac{1}{r} \mathbf{I}_r u_i\right) - (\mu^T \mathbf{0})^2 \\
&= \frac{1}{r} E \norm{u_i}_2^2 \\
&= \frac{1}{r} E \left[\tr(u_i u_i^T)\right] \\
&= \frac{1}{r} \tr\left(\Sigma + \mu \mu^T\right) \\
&= \frac{1}{r}\left(\tr \Sigma + \norm{\mu}_2^2\right)
\end{align*}

\bigskip
\FloatBarrier

\section{Active strategy in latent space} \label{sec:cces16-2d-viz}

\begin{figure}
\centering
\includegraphics[width=0.5\textwidth]{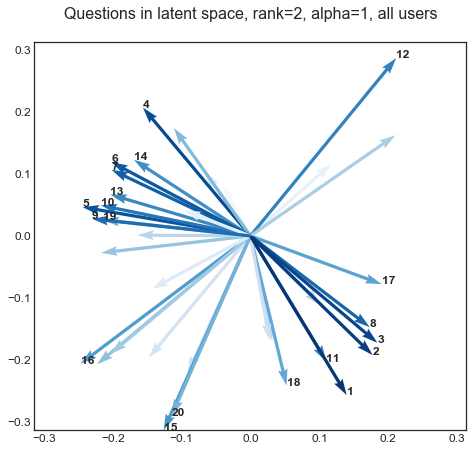}
\caption{We show the positions of the question factors in latent space using a rank-2 decomposition. The darker a vector is shaded, the earlier the corresponding question is asked by the active strategy. We label the first questions selected by the active strategy.}
\label{fig:cces16-position}
\end{figure}

To better understand how the active strategy operates, we visualize the relationship between question factors and question order for the 2016 CCES. We use a rank-2 decomposition for ease of visualization. Figure \ref{fig:cces16-position} depicts question factors as vectors in latent space. Longer vectors indicate questions that feature more prominently in the principal components. We use the term ``principal components'' loosely: recall SoftImpute is solved via soft-thresholded SVD. Unregularized SoftImpute performs principal component analysis on some complete version of the responses.

\begin{figure}[!htb]
\centering
\includegraphics[width=0.5\textwidth]{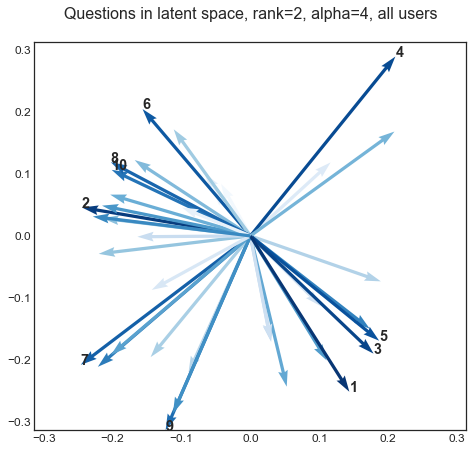}
\caption{Positions of the question factors in latent space using a rank-2 decomposition, CCES 2016. This time we set $\alpha=4$, increasing the precision and hence the information content of responses.}
\label{fig:cces16-position-alpha4}
\end{figure}

The active strategy front-loads medium-length vectors in the second and fourth quadrants -- the direction where the user factors have greatest prior variance. The active strategy spends its initial question budget sweeping around this direction of latent space. The longest vectors, which lie in the orthogonal direction, are asked midway through the survey. Short vectors are skipped, as they contribute little to precision. In simulations with higher $\alpha$, the relative information conveyed in each response is greater, and the longest vectors are chosen earlier (Figure \ref{fig:cces16-position-alpha4}).

\begin{figure}
  \centering
  \begin{subfigure}[b]{0.49\textwidth}
    \includegraphics[width=\textwidth]{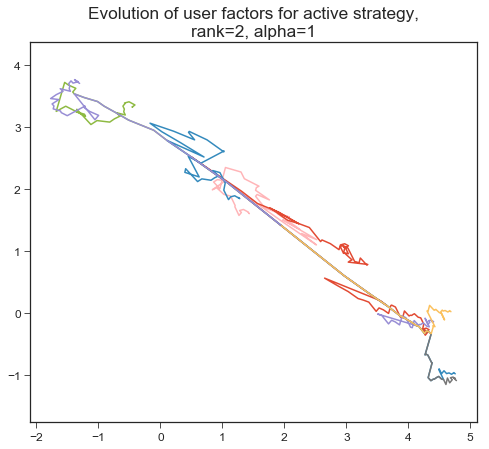}
    \caption{Posterior mean for 10 users over the survey, shown in different colors. All trajectories begin at the same prior mean.}
    \label{fig:cces16-user-mean}
  \end{subfigure}
  \hfill
  \begin{subfigure}[b]{0.49\textwidth}
    \includegraphics[width=\textwidth]{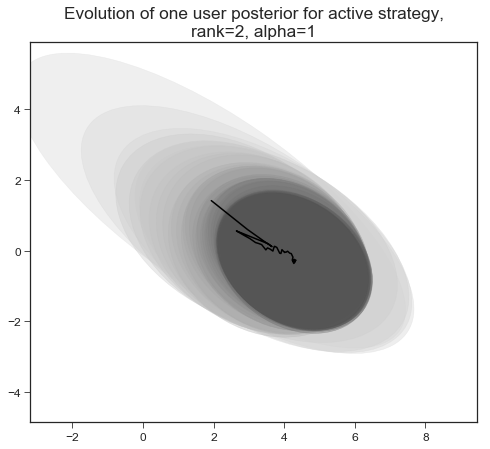}
    \caption{Posterior with 2$\sigma$ confidence region for a single user, starting light gray and darkening over the survey. The black line denotes the posterior mean.}
    \label{fig:cces16-user-posterior}
  \end{subfigure}
  \caption{We visualize the evolution of the user factors posterior over the course of the survey. As questions are asked, the posterior mean for a user moves in the direction of the corresponding question factors by increments depending on the response values and alpha. The confidence region also narrows in this direction, representing precision gained.}
\end{figure}

Visualizing the posterior of user factors during the survey reinforces this story. Sample user trajectories in Figure \ref{fig:cces16-user-mean} suggest that the initial questions in the active ordering lie in the direction of greatest user variance. The active ordering first places users along this direction before seeking information along the orthogonal direction. Most precision gains occur along this direction, early in the survey, as the narrowing confidence region in Figure \ref{fig:cces16-user-posterior} shows. The confidence region quantifies the declining uncertainty in user position as more questions are asked.

\FloatBarrier

\section{Additional results for 2016 CCES}

\begin{figure}[!htb]
\centering
\includegraphics[scale=0.4]{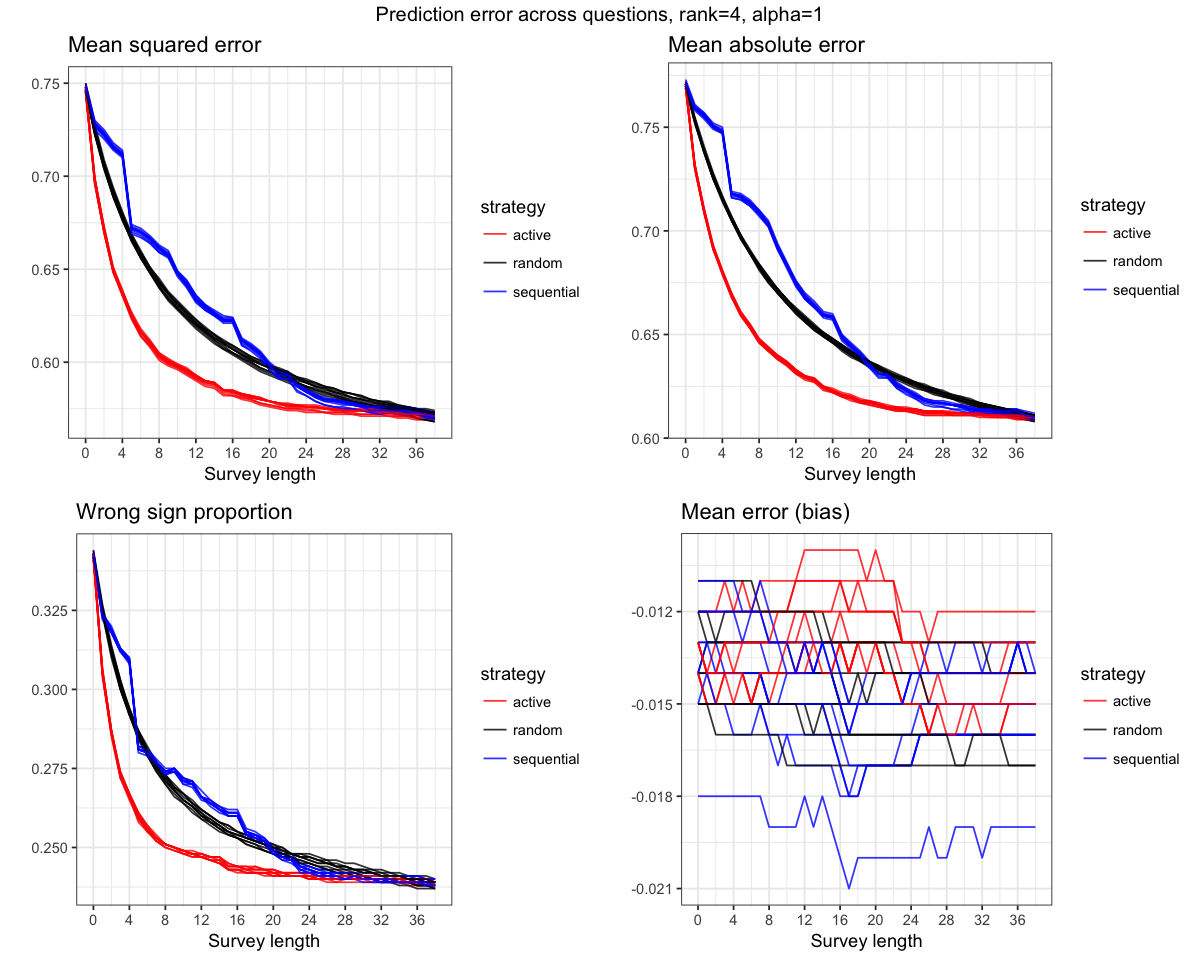}
\caption{Prediction error on the sparse holdout set, measured across 10 iterations of simulating the 2016 CCES. Error metrics include mean squared error, mean absolute error, the proportion of predictions with the wrong sign, and mean signed error or bias. Active question selection attains lower mean squared and mean absolute prediction error than baselines. The active strategy also correctly predicts the sign of survey responses more often. \label{fig:cces16-compare-metrics-extra}}
\end{figure}

\begin{figure}[!htb]
\centering
\includegraphics[width=\textwidth]{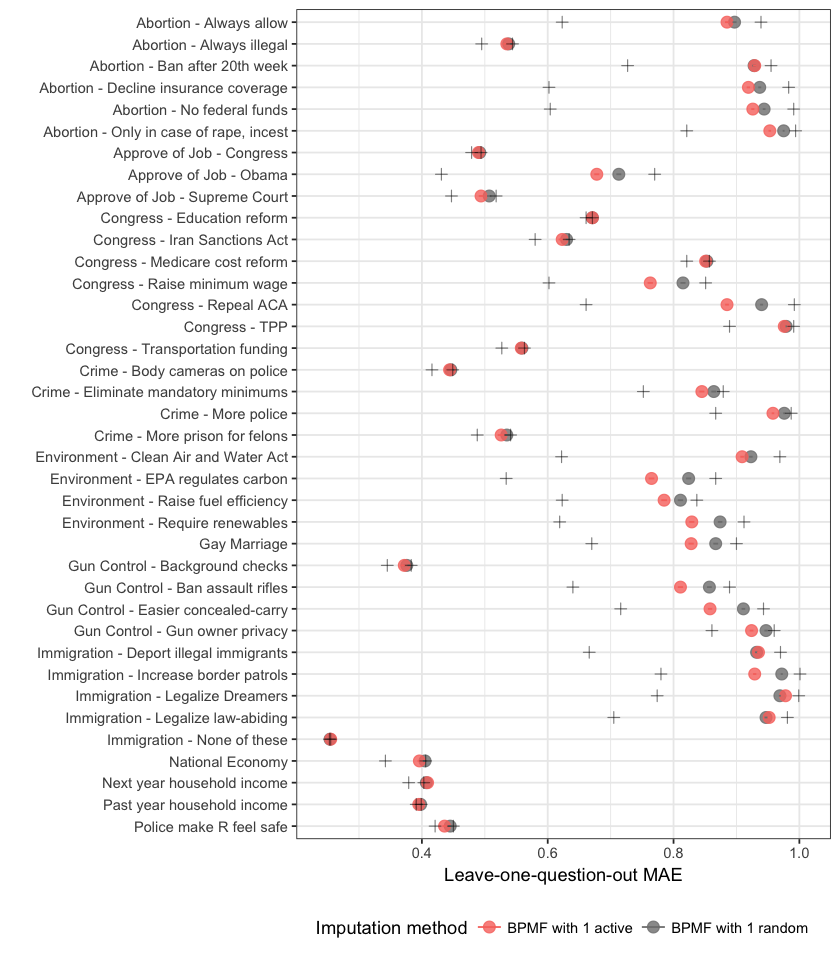}
\caption{Mean absolute prediction error per question after one question chosen actively or randomly, CCES 2016.}
\label{fig:cces16-compare-questions-1}
\end{figure}

\begin{figure}[!htb]
\centering
\includegraphics[width=\textwidth]{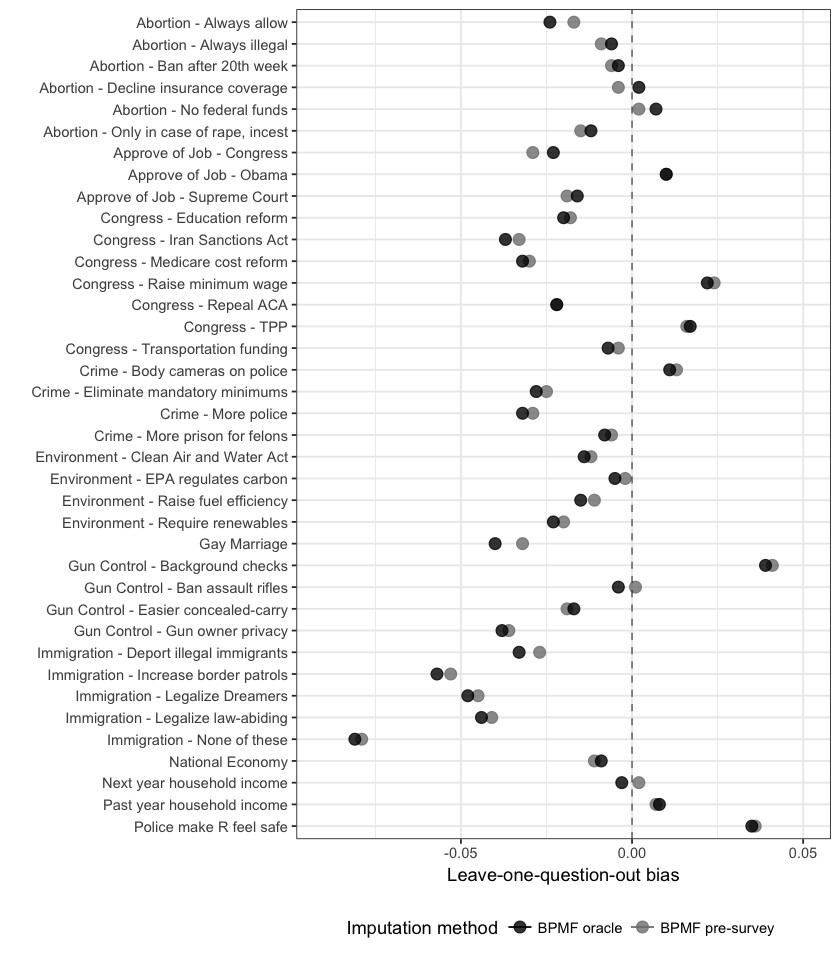}
\caption{Bias per question when imputing responses with no knowledge (``pre-survey'') and with all other responses revealed (``oracle'').}
\label{fig:cces16-compare-questions-bias}
\end{figure}

\begin{figure}[!htb]
\centering
\includegraphics[width=0.5\textwidth]{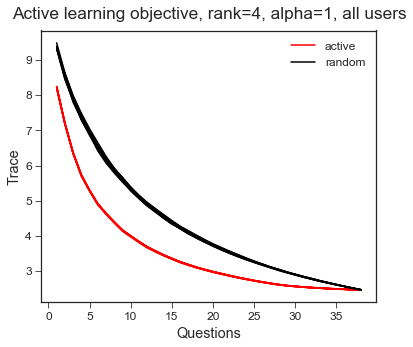}
\caption{We plot the trace of posterior variance achieved by 10 simulations of the active and random strategies on 2016 CCES. Note the active strategy produces a single question ordering across all users, and thus a single sequence of objective function values per simulation. This sequence varies slightly across simulations due to randomness in estimated question factors. To compute the objective function for the random strategy in one simulation, we average the trace of posterior variance after each question across 100 random question orderings.}
\label{fig:cces16-objective}
\end{figure}

\begin{figure}
\centering
\includegraphics[width=\textwidth]{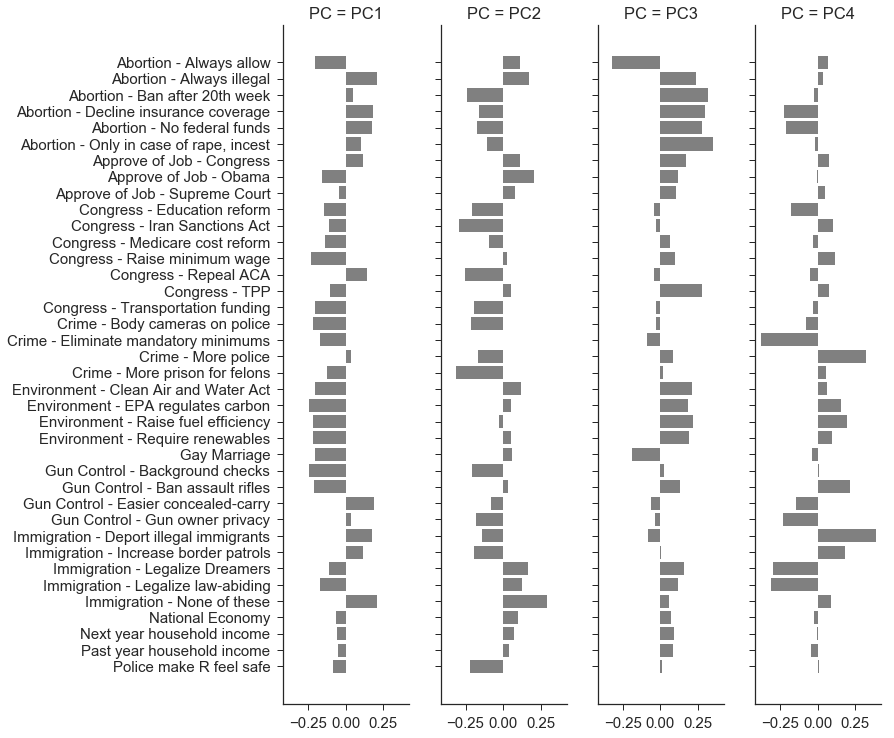}
\caption{Question factors for 2016 CCES using a rank-4 matrix decomposition.}
\label{fig:cces16-pcs}
\end{figure}

\begin{figure}
  \centering
  \begin{subfigure}[b]{0.55\textwidth}
    \includegraphics[width=\textwidth]{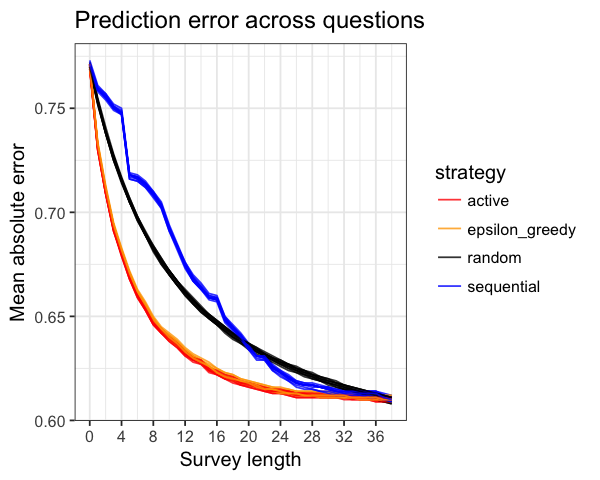}
    \caption{Prediction error on the sparse holdout set, measured across 10 simulations.}
  \end{subfigure}
  \hfill
  \begin{subfigure}[b]{0.40\textwidth}
    \includegraphics[width=\textwidth]{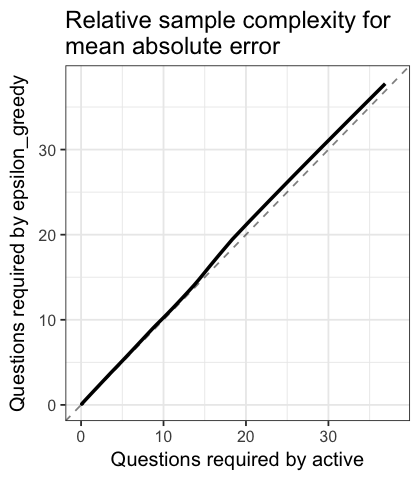}
    \caption{Relative sample complexity of the active and $\epsilon$-greedy strategies.}
  \end{subfigure}
  \caption{We add to the overall error comparison the $\epsilon$-greedy strategy, which selects a random question with probability 0.05 and otherwise follows the active order. $\epsilon$-greedy question selection is no better than active for PMF with fixed question factors. \label{fig:cces16-epsilon-greedy-compare-metrics}}
\end{figure}

\begin{figure}
  \centering
  \begin{subfigure}[b]{0.55\textwidth}
    \includegraphics[width=\textwidth]{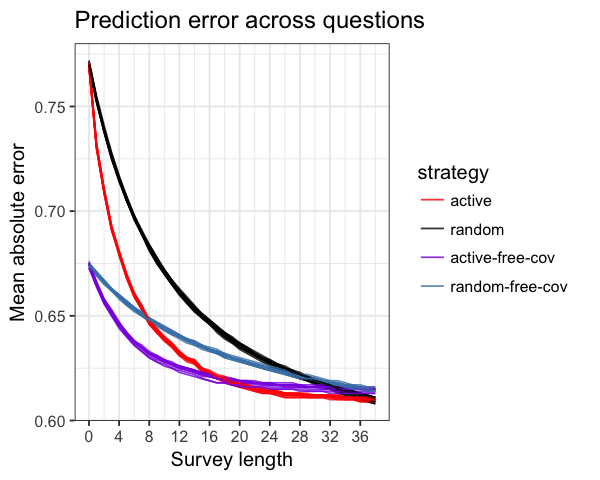}
    \caption{Prediction error on the sparse holdout set, measured across 10 simulations.}
  \end{subfigure}
  \hfill
  \begin{subfigure}[b]{0.40\textwidth}
    \includegraphics[width=\textwidth]{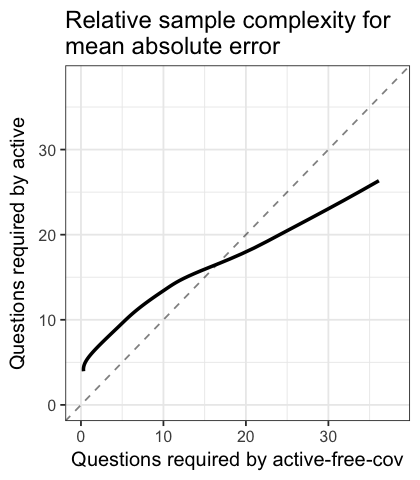}
    \caption{Relative complexity of the active strategy with and without free covariates.}
  \end{subfigure}
  \caption{We plot overall imputation error when responses to covariate questions are always available (``free-cov''). For short simulated surveys, information from free covariates narrows down user position in latent space, reducing error faster for both active and random strategies. As survey length grows, there is a strategy-specific crossover point after which using free covariates leads to slightly higher prediction error. This occurs because the free covariates participate in matrix factorization: the loss function includes terms for their reconstruction error. This changes the question factors estimated from the training half. Meanwhile, imputation error is evaluated on held-out survey responses only, not covariates. We have biased question factors away from those that would optimize the original covariate-free loss function, in exchange for covariate-enabled variance reduction. \label{fig:cces16-free-covariates-compare-metrics}}
\end{figure}

\FloatBarrier

\section{Tuning simulation parameters} \label{sec:tune-sim-params}

\begin{figure}[!htb]
\centering
\includegraphics[width=0.8\textwidth]{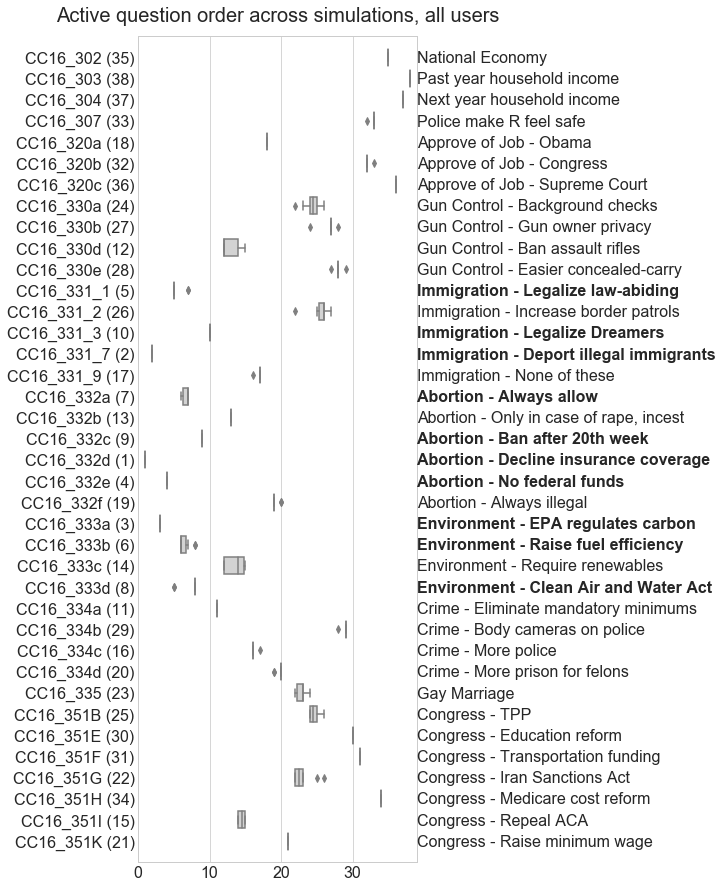}
\caption{Active ordering for 2016 CCES using a rank-4 matrix decomposition and D-optimality. \label{fig:cces16-question-rank-d-optimality}}
\end{figure}

\begin{figure}[!htb]
\centering
\includegraphics[width=0.8\textwidth]{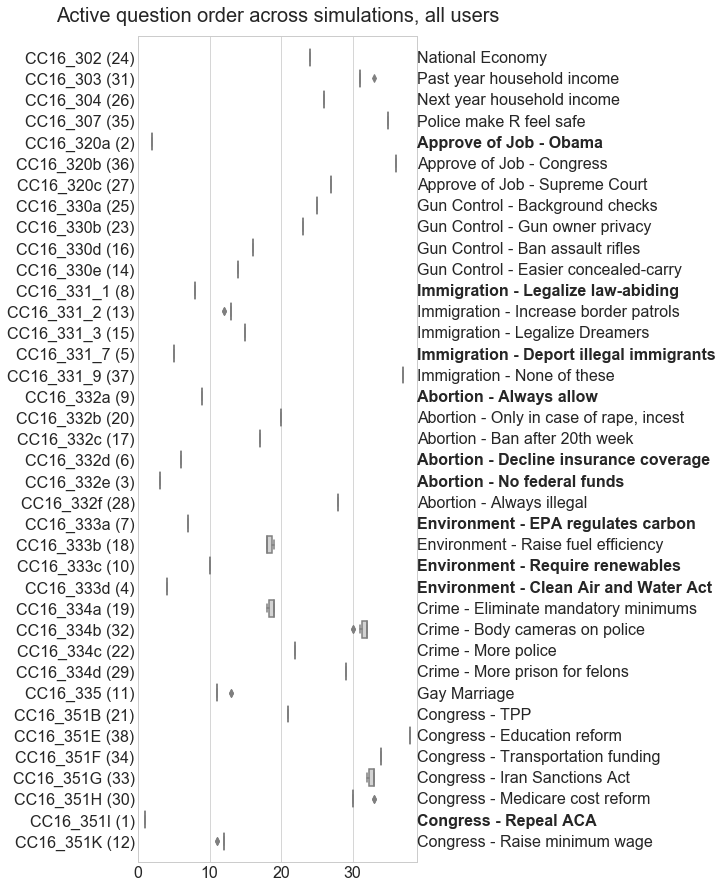}
\caption{Active ordering for 2016 CCES using a rank-4 matrix decomposition and E-optimality. \label{fig:cces16-question-rank-e-optimality}}
\end{figure}

\begin{figure}[!htb]
  \centering
  \begin{subfigure}[b]{0.48\textwidth}
    \includegraphics[width=\textwidth]{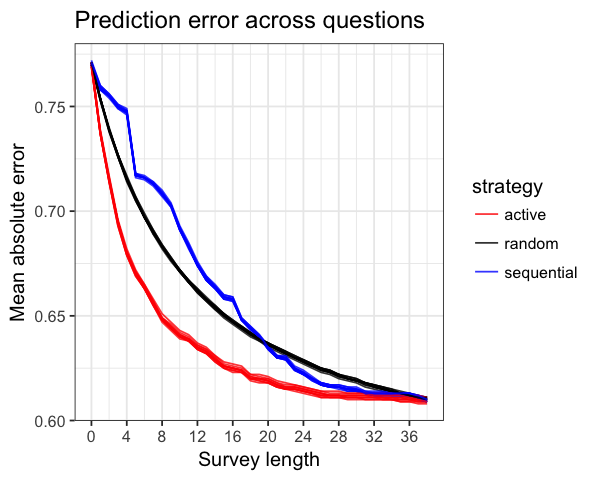}
    \caption{Minimizing the determinant of posterior variance (D-optimality).}
  \end{subfigure}
  \hfill
  \begin{subfigure}[b]{0.48\textwidth}
    \includegraphics[width=\textwidth]{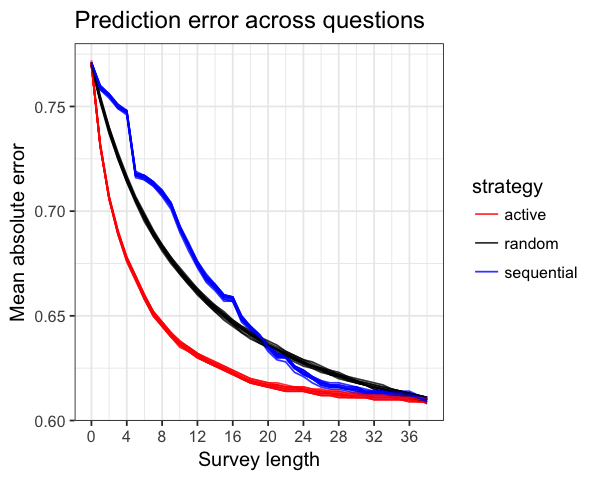}
    \caption{Minimizing the maximum eigenvalue of posterior variance (E-optimality).}
  \end{subfigure}
  \caption{Prediction error on the sparse holdout set across 10 simulations of the 2016 CCES. Active question selection uses different optimality criteria.}
  \label{fig:cces16-compare-metrics-de-optimality}
\end{figure}

\begin{figure}[!htb]
\centering
\includegraphics[width=0.8\textwidth]{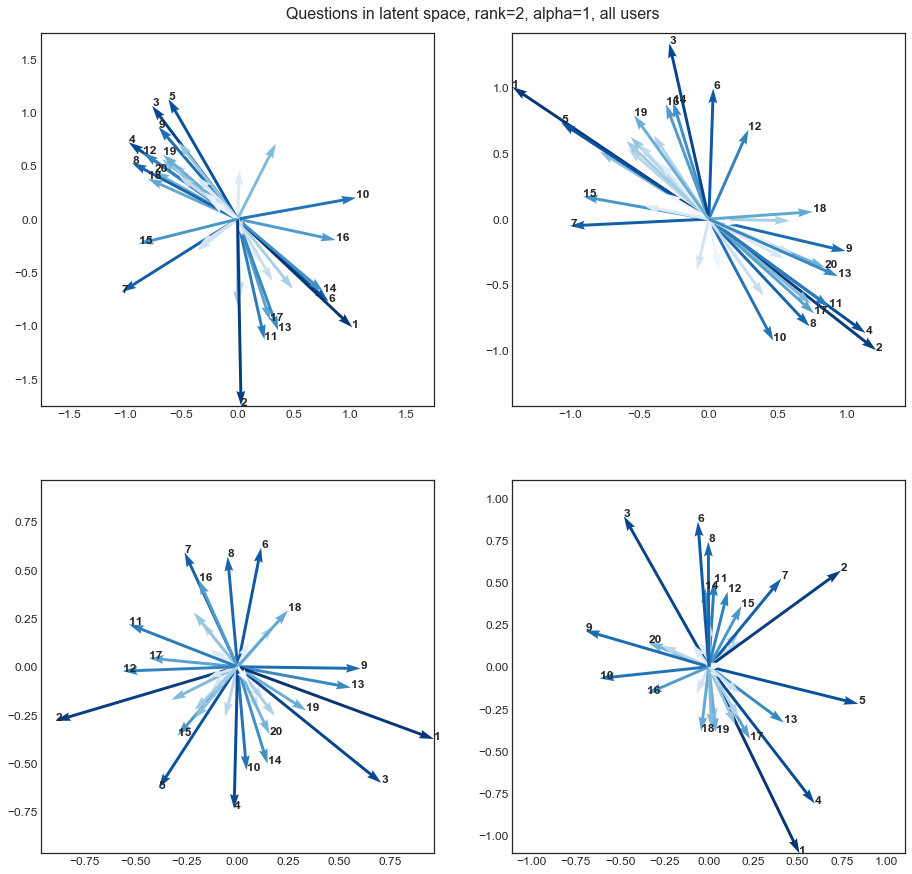}
\caption{Positions of the question factors in latent space, estimated with Frobenius norm regularization instead of SoftImpute, for different simulations of CCES 2016.}
\label{fig:cces16-position-frobenius}
\end{figure}

\begin{figure}[!htb]
\centering
\includegraphics[width=0.8\textwidth]{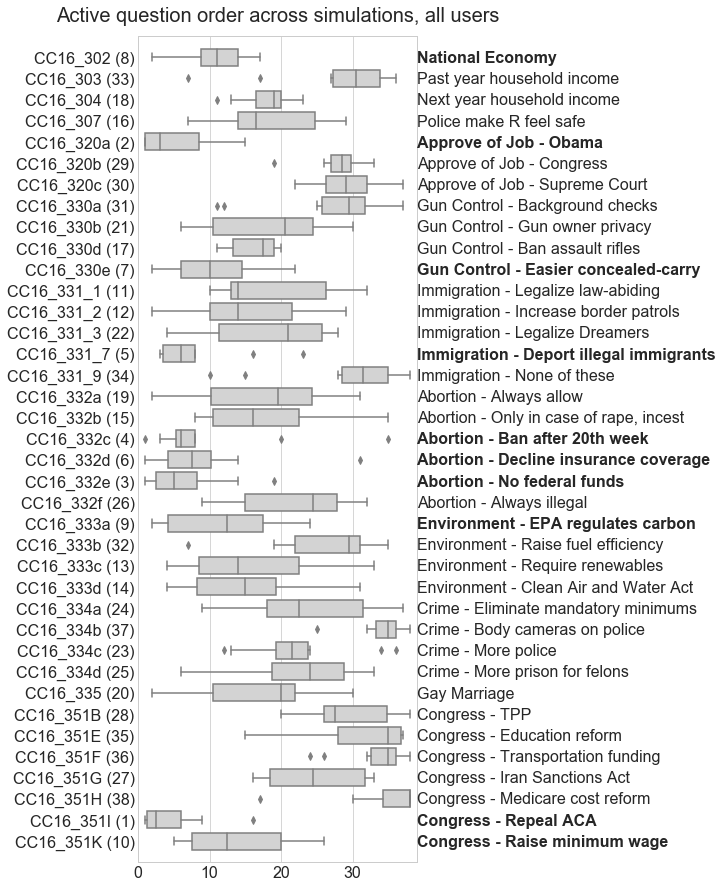}
\caption{Active ordering for 2016 CCES using a rank-4 matrix decomposition. Question factors are estimated with Frobenius norm regularization instead of SoftImpute.}
\label{fig:cces16-question-rank-frobenius}
\end{figure}

\begin{figure}[!htb]
\centering
\includegraphics[width=0.5\textwidth]{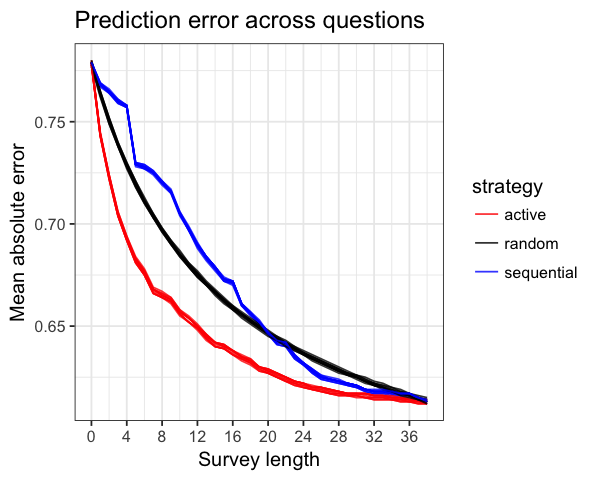}
\caption{Prediction error on the sparse holdout set across 10 simulations of the 2016 CCES using a rank-8 matrix decomposition. This achieves roughly the same final imputation error as the rank-4 decomposition (Figure \ref{fig:cces16-compare-metrics}). Moreover, overall error with $r=8$ declines more slowly as responses are revealed. For the active strategy, 8 questions with $r=4$ give the same MAE as 12 questions with $r=8$. For the random strategy, 6 questions with $r=4$ give the same MAE as 8 questions with $r=8$.}
\label{fig:cces16-compare-metrics-rank8}
\end{figure}

\FloatBarrier

\section{Robustness checks for question order} \label{sec:question-order-robustness}

\begin{figure}[!htb]
\centering
\includegraphics[width=0.8\textwidth]{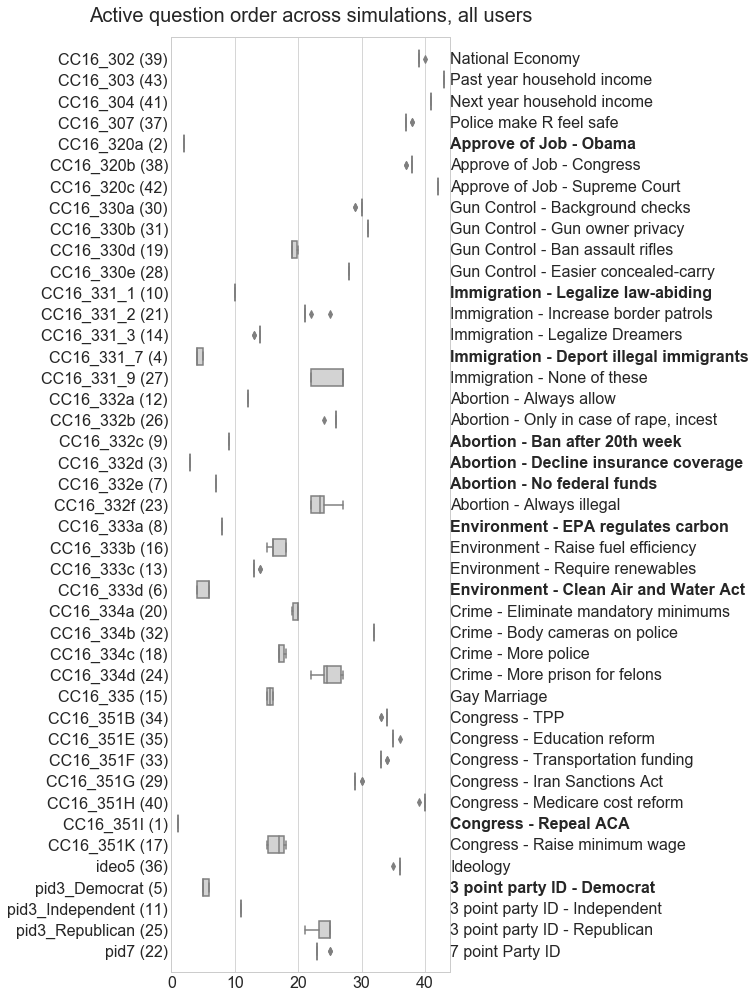}
\caption{Question rank across 10 simulations of the active strategy on the 2016 CCES. This iteration adds questions about political affiliation. Of these questions, the most informative is identifying as Democrat, followed by identifying as independent. Identifying as Republican and rating one's ideology on an ordinal scale do not appear in the top 20. \label{fig:cces16-question-rank-with-pol-ids}}
\end{figure}

\begin{figure}[!htb]
\centering
\includegraphics[width=0.8\textwidth]{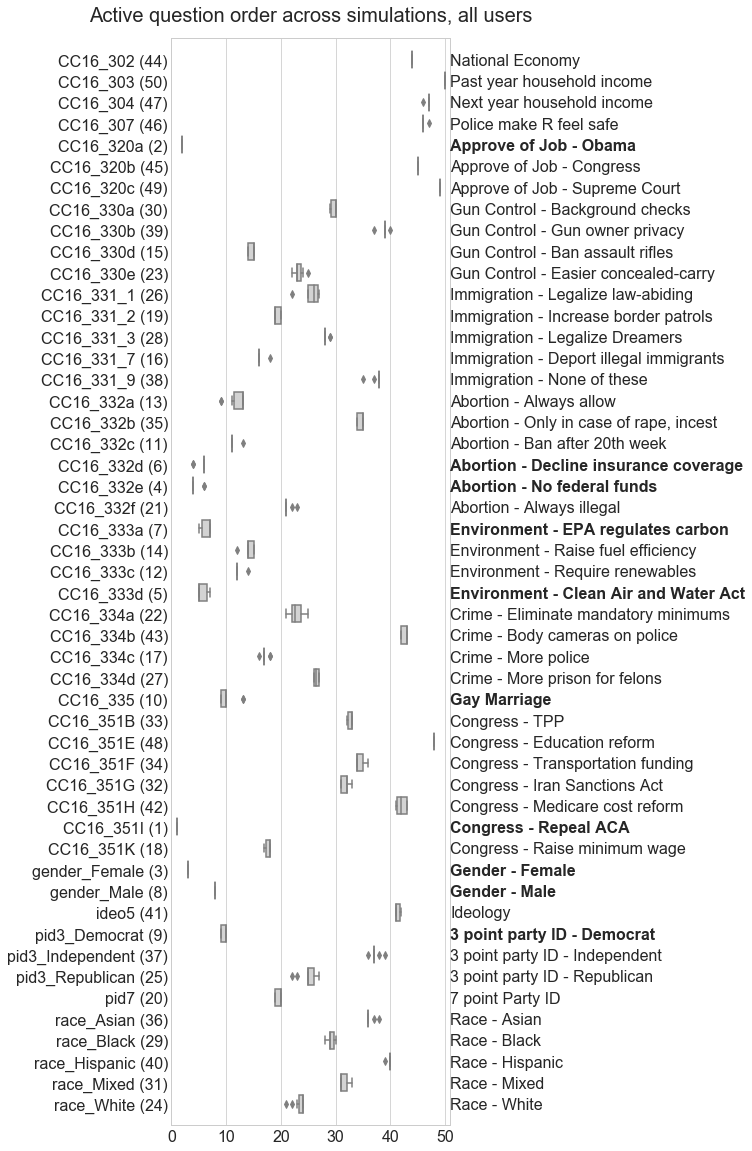}
\caption{Question rank across 10 simulations of the active strategy on the 2016 CCES. This iteration adds questions about political affiliation and demographics. Gender questions are asked early, as is identifying as Democrat. Questions about race do not appear in the top 20. Immigration questions and identifying as independent become less important. \label{fig:cces16-question-rank-with-demos}}
\end{figure}

\begin{figure}[!htb]
\centering
\includegraphics[width=0.8\textwidth]{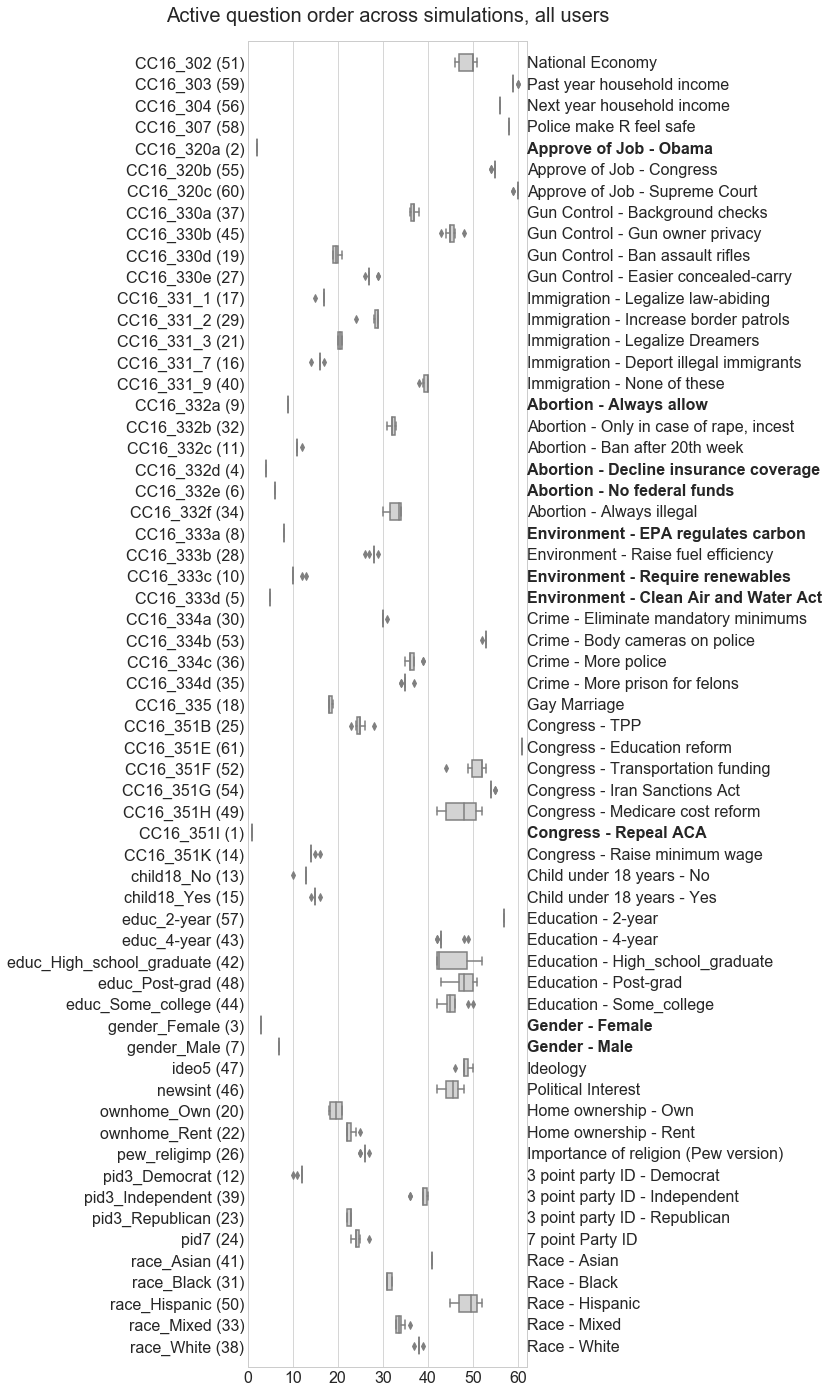}
\caption{Question rank across 10 simulations of the active strategy on the 2016 CCES. This iteration adds questions about political affiliation, demographics, education, financial well-being, and religiosity. Of the new questions, child and home ownership are prioritized, but only after gender and Democrat self-identification. \label{fig:cces16-question-rank-full}}
\end{figure}

\begin{figure}[!htb]
\centering
\includegraphics[width=0.8\textwidth]{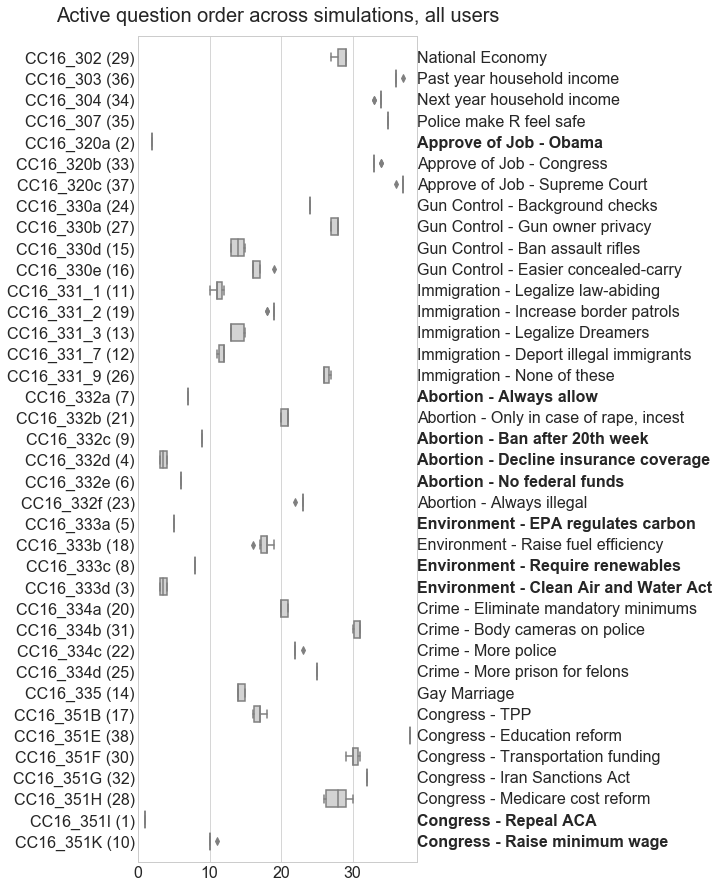}
\caption{Question rank across 10 simulations of the active strategy on the 2016 CCES. All covariate questions are automatically included at the start of the survey; the active ordering takes information from those questions into account. \label{fig:cces16-question-rank-free-covariates}}
\end{figure}

\FloatBarrier

\section{Results for 2018 CCES}  \label{sec:cces18-results}

\begin{figure}[!htb]
  \centering
  \begin{subfigure}[b]{0.55\textwidth}
    \includegraphics[width=\textwidth]{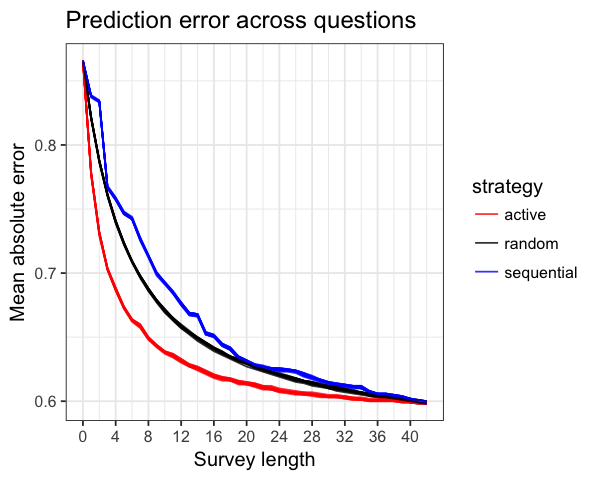}
    \caption{Prediction error on the sparse holdout set, measured across 10 simulations.}
    \label{fig:cces18-compare-metrics}
  \end{subfigure}
  \hfill
  \begin{subfigure}[b]{0.40\textwidth}
    \includegraphics[width=\textwidth]{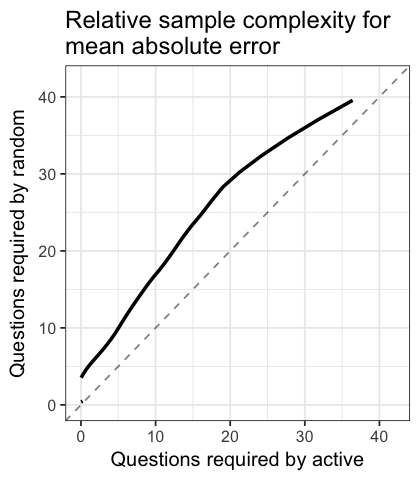}
    \caption{Sample complexity of the active strategy relative to random-order questions.}
    \label{fig:cces18-relative-complexity}
  \end{subfigure}
  \caption{Summary measures of imputation error for simulating the 2018 CCES using each question selection strategy.}
\end{figure}


\begin{figure}[!htb]
  \centering
  \begin{subfigure}{0.48\textwidth}
    \includegraphics[width=\textwidth]{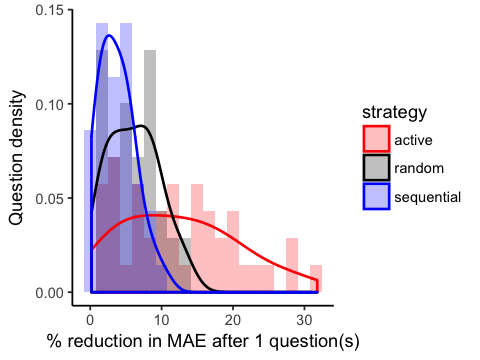}
  \end{subfigure}
  \hfill
  \begin{subfigure}{0.48\textwidth}
    \includegraphics[width=\textwidth]{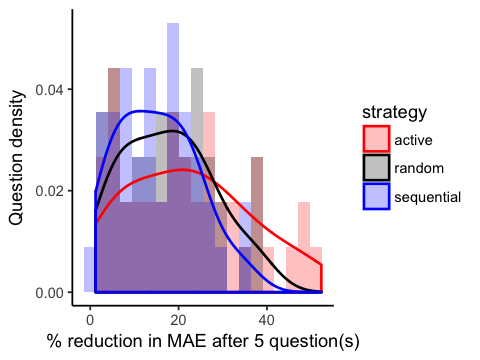}
  \end{subfigure}
  \caption{Reduction in imputation error from pre-survey levels for the 2018 CCES.}
  \label{fig:cces18-error-reduction}
\end{figure}

\begin{figure}[!htb]
\centering
\includegraphics[width=\textwidth]{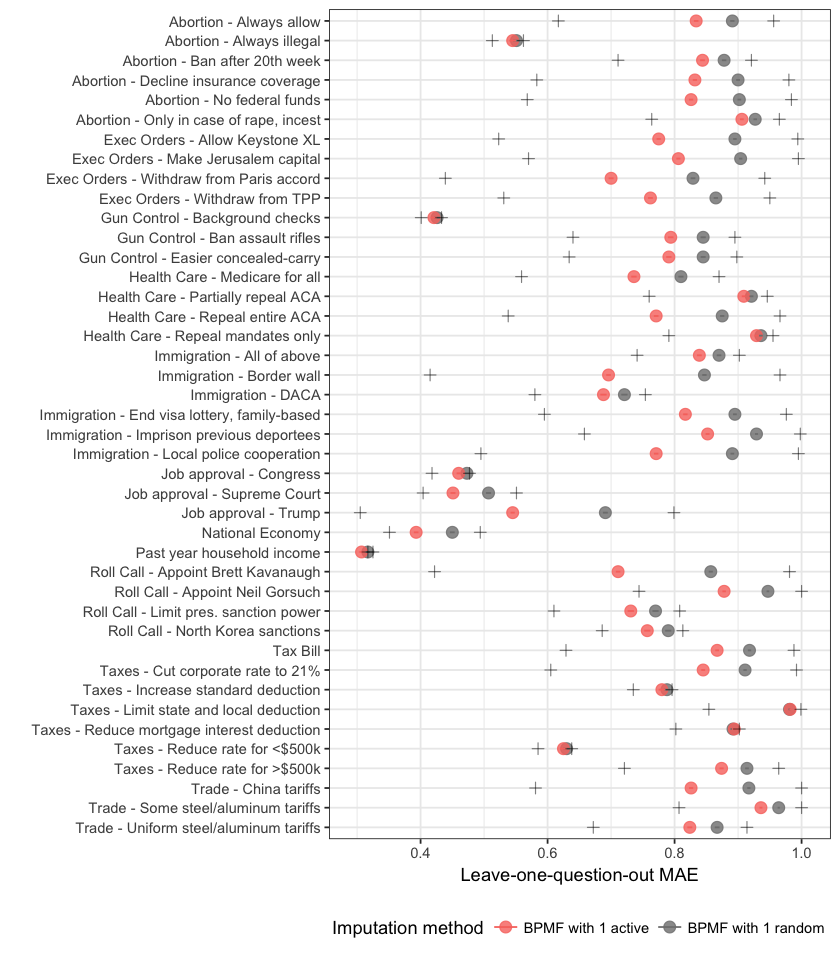}
\caption{Mean absolute prediction error per question after one question chosen actively or randomly, CCES 2018.}
\label{fig:cces18-compare-questions-1}
\end{figure}

\begin{figure}[!htb]
\centering
\includegraphics[width=\textwidth]{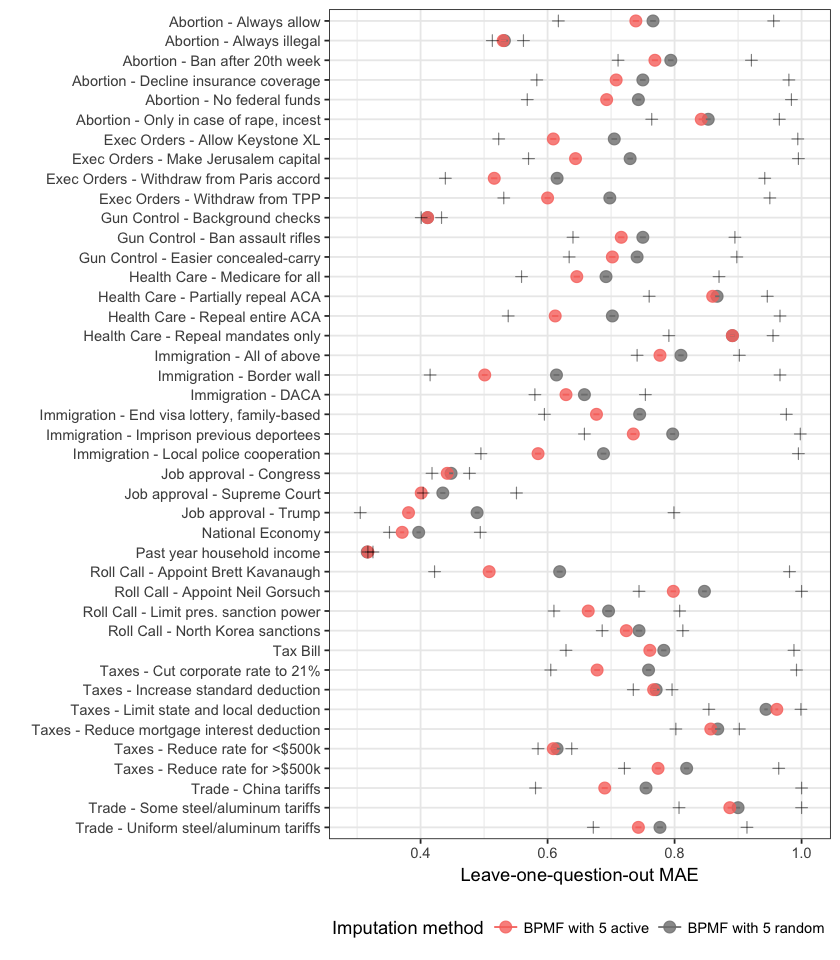}
\caption{Mean absolute prediction error per question after five questions chosen actively or randomly, CCES 2018.}
\label{fig:cces18-compare-questions-5}
\end{figure}

\begin{figure}[!htb]
\centering
\includegraphics[width=\textwidth]{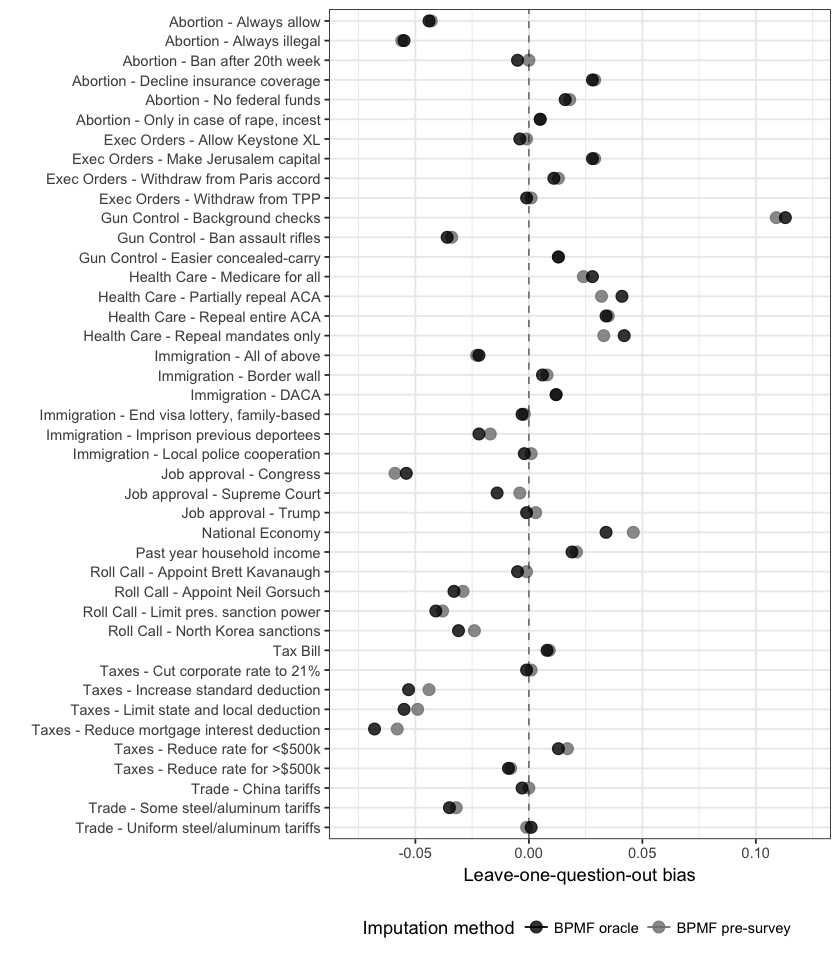}
\caption{Bias per question when imputing responses with no knowledge (``pre-survey'') and with all other responses revealed (``oracle''), CCES 2018.}
\label{fig:cces18-compare-questions-bias}
\end{figure}

\begin{figure}[!htb]
\centering
\includegraphics[width=0.8\textwidth]{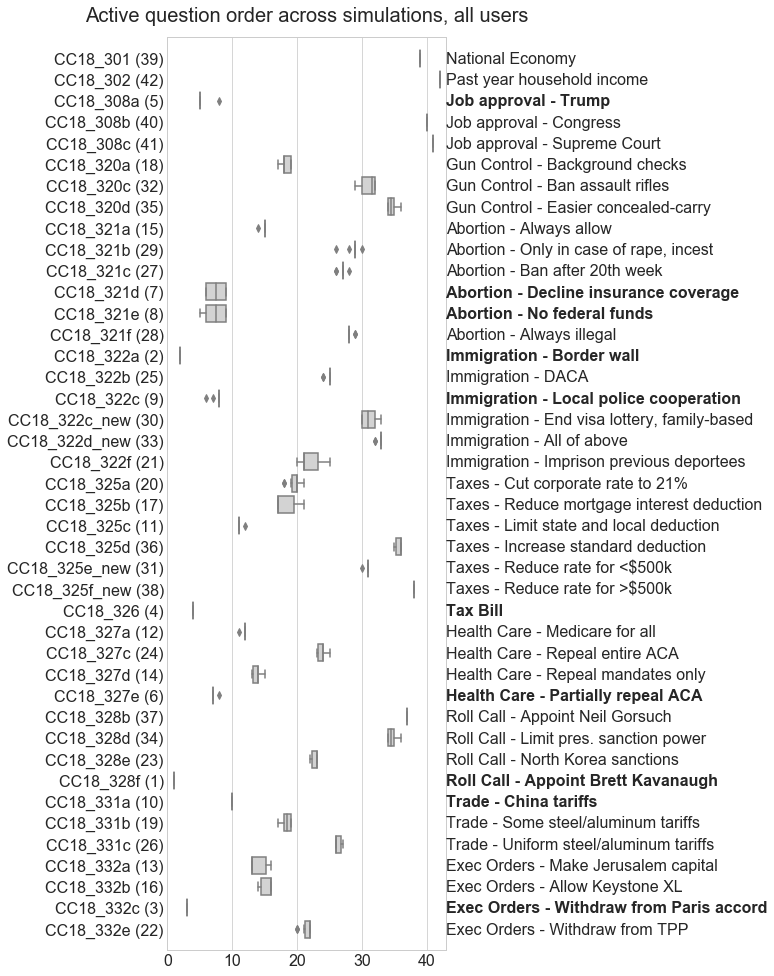}
\caption{Active ordering for 2018 CCES using a rank-4 matrix decomposition. \label{fig:cces18-question-rank}}
\end{figure}

\begin{figure}[!htb]
\centering
\includegraphics[width=0.5\textwidth]{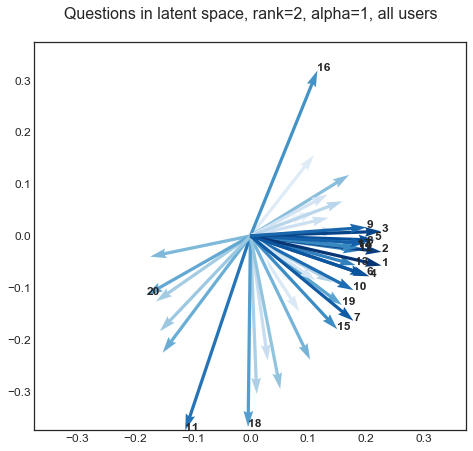}
\caption{Positions of the question factors in latent space using a rank-2 decomposition, CCES 2018.}
\label{fig:cces18-position}
\end{figure}

\begin{figure}[!htb]
\centering
\includegraphics[width=\textwidth]{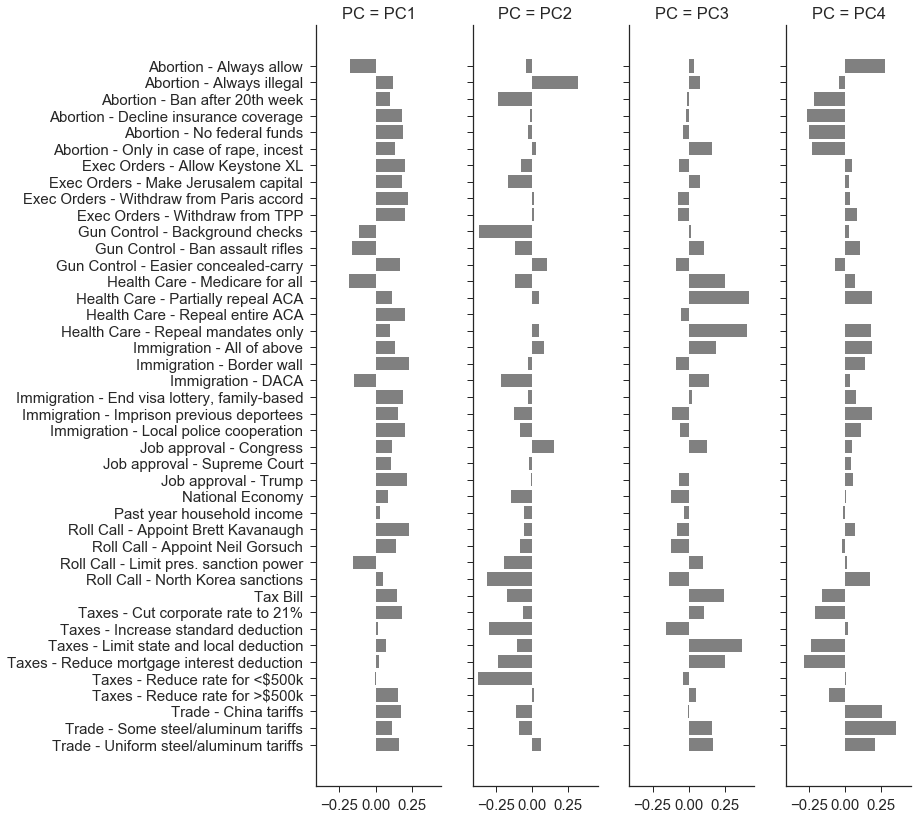}
\caption{Question factors for 2018 CCES using a rank-4 matrix decomposition.}
\label{fig:cces18-pcs}
\end{figure}

\FloatBarrier

\section{Results for 2012 CCES}  \label{sec:cces12-results}

\begin{figure}[!htb]
  \centering
  \begin{subfigure}[b]{0.55\textwidth}
    \includegraphics[width=\textwidth]{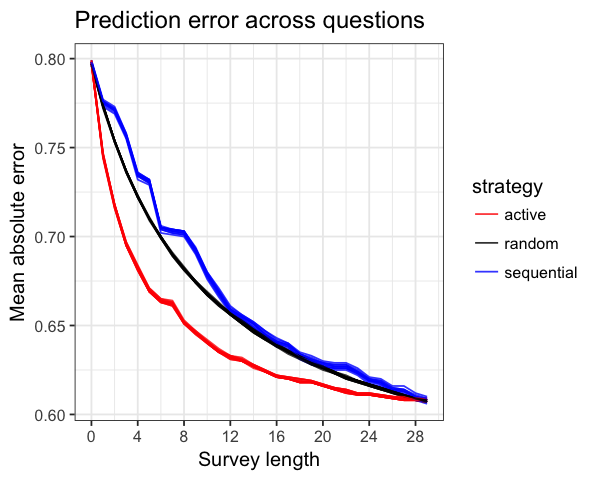}
    \caption{Prediction error on the sparse holdout set, measured across 10 simulations.}
    \label{fig:cces12-compare-metrics}
  \end{subfigure}
  \hfill
  \begin{subfigure}[b]{0.40\textwidth}
    \includegraphics[width=\textwidth]{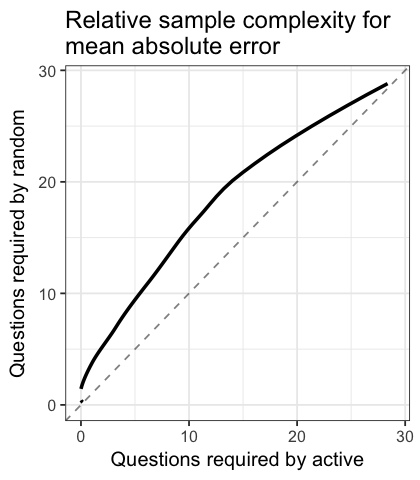}
    \caption{Sample complexity of the active strategy relative to random-order questions.}
    \label{fig:cces12-relative-complexity}
  \end{subfigure}
  \caption{Summary measures of imputation error for simulating the 2012 CCES using each question selection strategy.}
\end{figure}


\begin{figure}[!htb]
  \centering
  \begin{subfigure}{0.48\textwidth}
    \includegraphics[width=\textwidth]{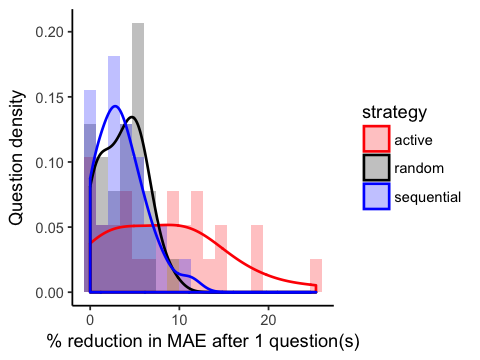}
  \end{subfigure}
  \hfill
  \begin{subfigure}{0.48\textwidth}
    \includegraphics[width=\textwidth]{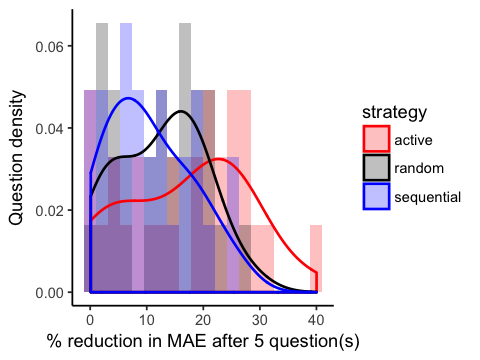}
  \end{subfigure}
  \caption{Reduction in imputation error from pre-survey levels for the 2012 CCES.}
  \label{fig:cces12-error-reduction}
\end{figure}

\begin{figure}[!htb]
\centering
\includegraphics[width=\textwidth]{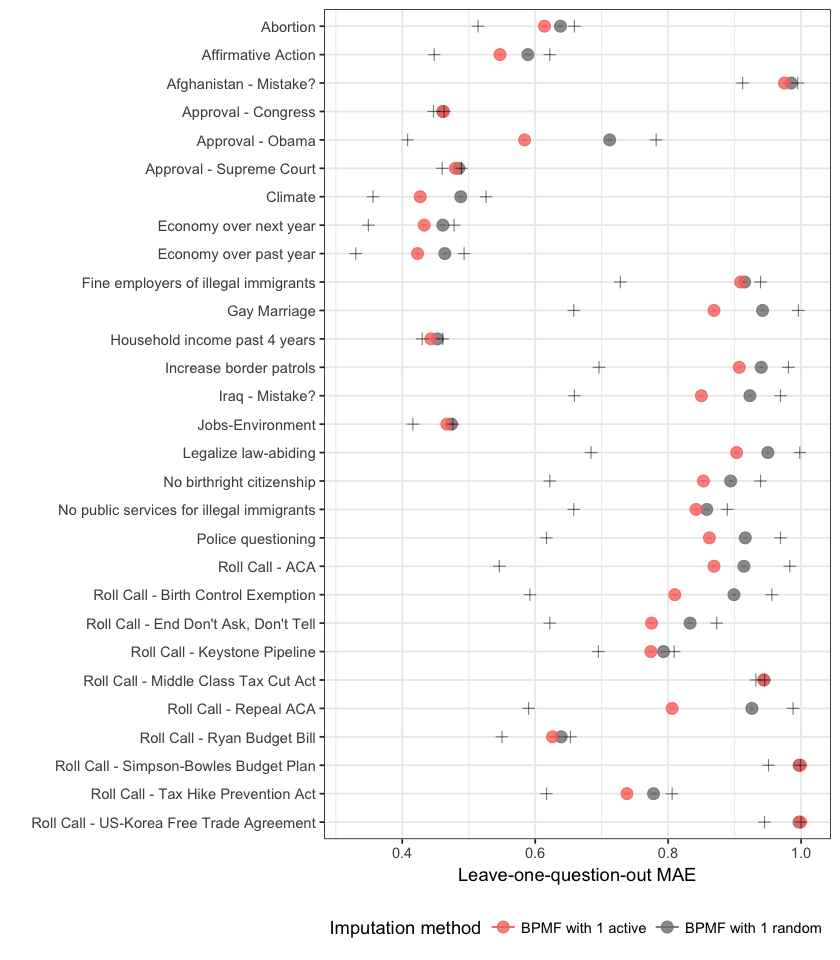}
\caption{Mean absolute prediction error per question after one question chosen actively or randomly, CCES 2012.}
\label{fig:cces12-compare-questions-1}
\end{figure}

\begin{figure}[!htb]
\centering
\includegraphics[width=\textwidth]{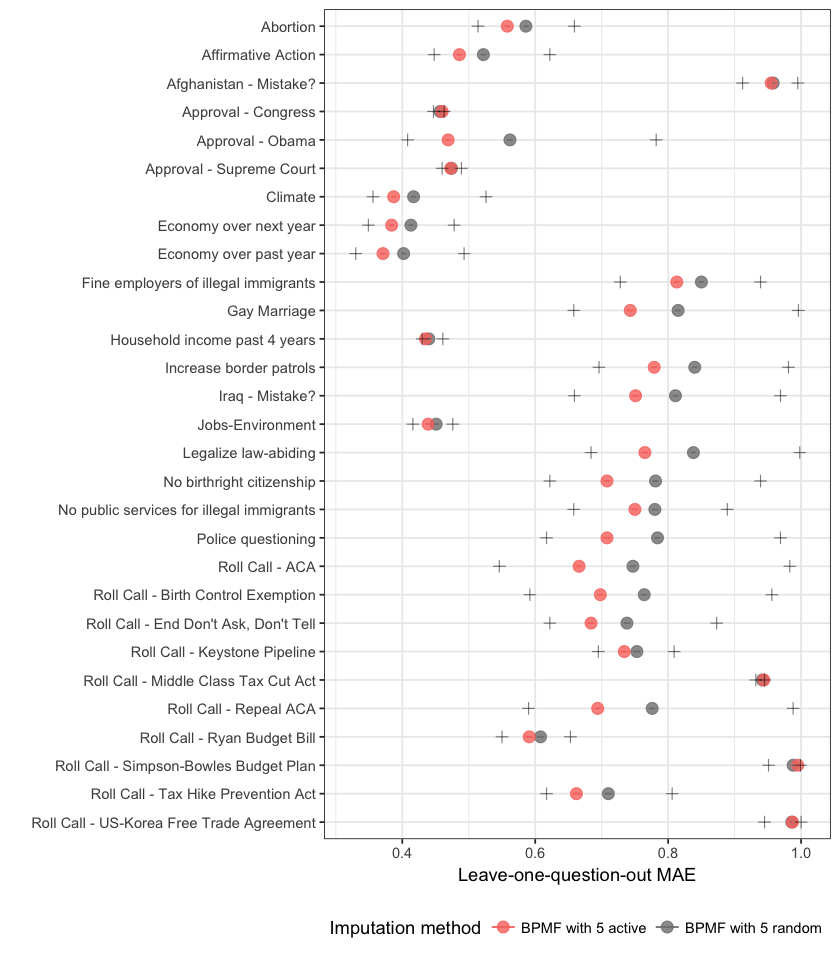}
\caption{Mean absolute prediction error per question after five questions chosen actively or randomly, CCES 2012.}
\label{fig:cces12-compare-questions-5}
\end{figure}

\begin{figure}[!htb]
\centering
\includegraphics[width=\textwidth]{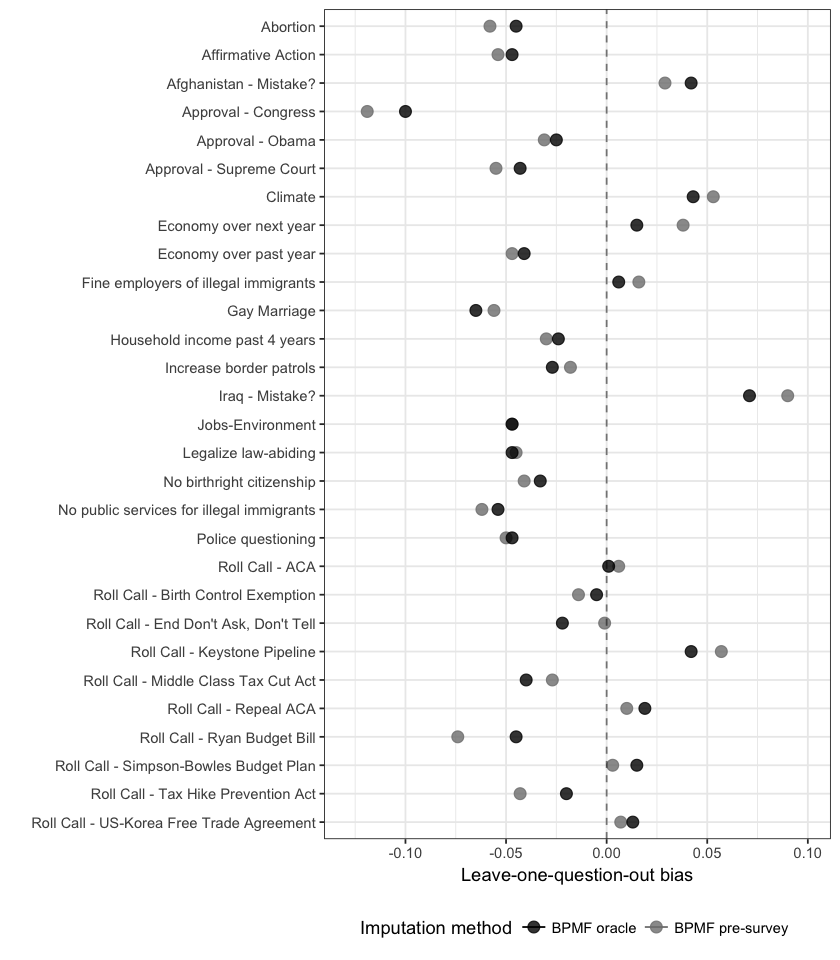}
\caption{Bias per question when imputing responses with no knowledge (``pre-survey'') and with all other responses revealed (``oracle''), CCES 2012.}
\label{fig:cces12-compare-questions-bias}
\end{figure}

\begin{figure}[!htb]
\centering
\includegraphics[width=0.8\textwidth]{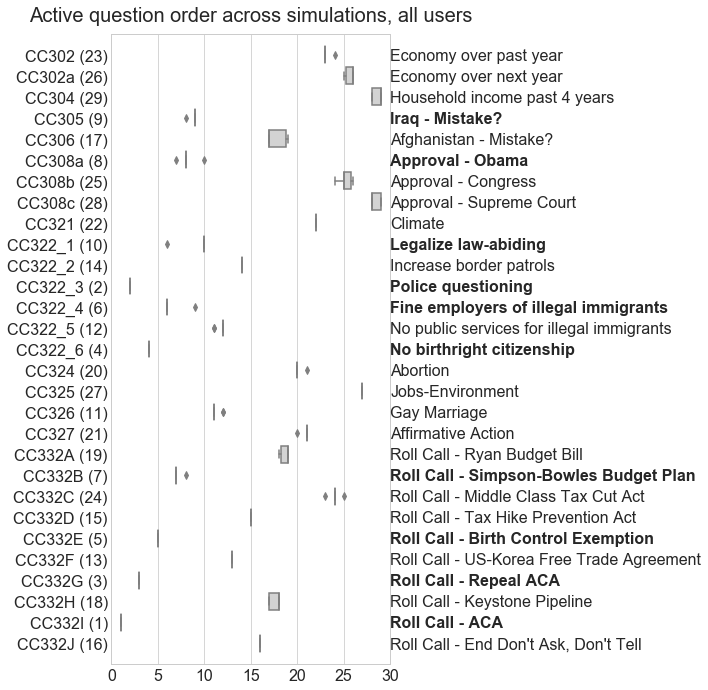}
\caption{Active ordering for 2012 CCES using a rank-4 matrix decomposition. \label{fig:cces12-question-rank}}
\end{figure}

\begin{figure}[!htb]
\centering
\includegraphics[width=0.5\textwidth]{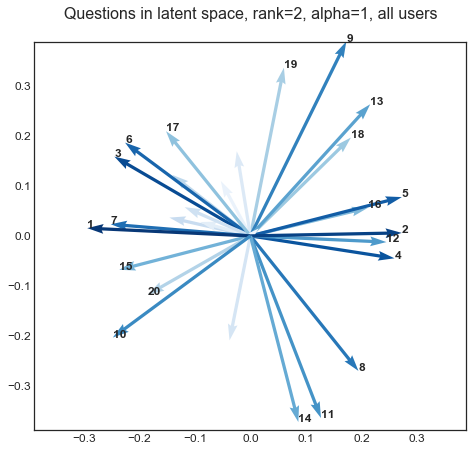}
\caption{Positions of the question factors in latent space using a rank-2 decomposition, CCES 2012.}
\label{fig:cces12-position}
\end{figure}

\begin{figure}[!htb]
\centering
\includegraphics[width=\textwidth]{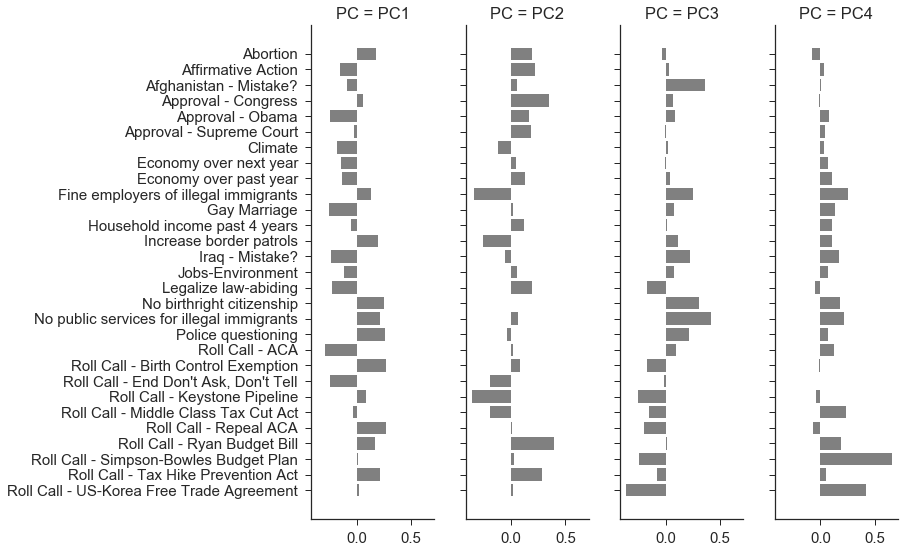}
\caption{Question factors for 2012 CCES using a rank-4 matrix decomposition.}
\label{fig:cces12-pcs}
\end{figure}

\FloatBarrier

\section{Results for first Facebook survey} \label{sec:facebook-results}

\begin{figure}[!htb]
  \centering
  \begin{subfigure}[b]{0.55\textwidth}
    \includegraphics[width=\textwidth]{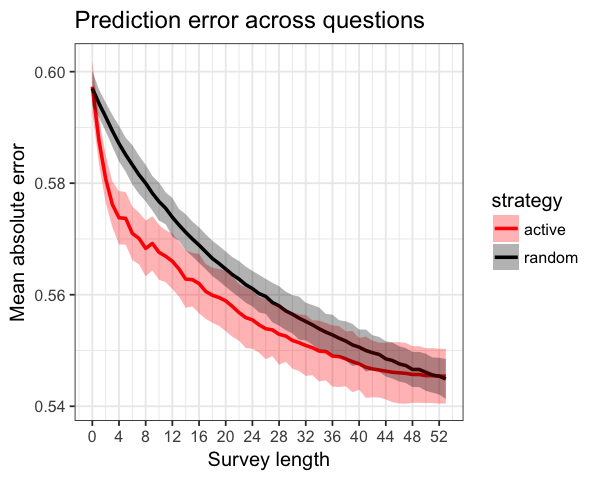}
    \caption{Prediction error on the sparse holdout set, measured across 10 simulations.}
    \label{fig:facebook-compare-metrics}
  \end{subfigure}
  \hfill
  \begin{subfigure}[b]{0.40\textwidth}
    \includegraphics[width=\textwidth]{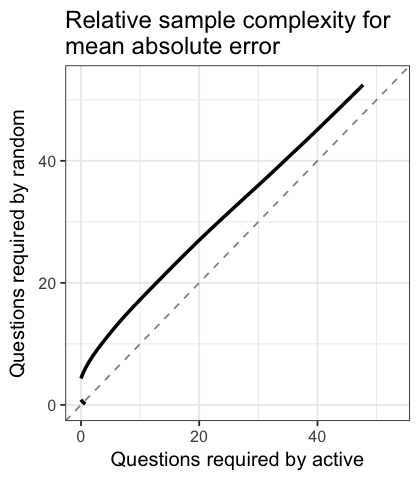}
    \caption{Sample complexity of the active strategy relative to random-order questions.}
    \label{fig:facebook-relative-complexity}
  \end{subfigure}
  \caption{Summary measures of imputation error for simulating the Facebook survey using each question selection strategy. For clarity, we plot $2\sigma$ uncertainty bands across 10 simulations of each strategy.}
\end{figure}


\begin{figure}[!htb]
  \centering
  \begin{subfigure}{0.48\textwidth}
    \includegraphics[width=\textwidth]{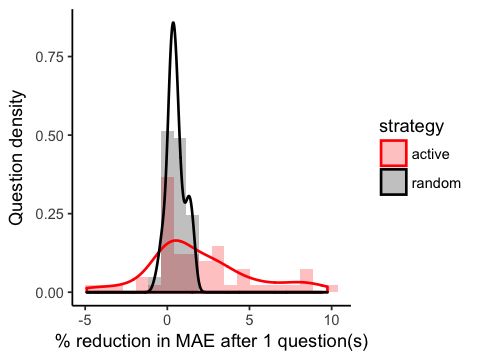}
  \end{subfigure}
  \hfill
  \begin{subfigure}{0.48\textwidth}
    \includegraphics[width=\textwidth]{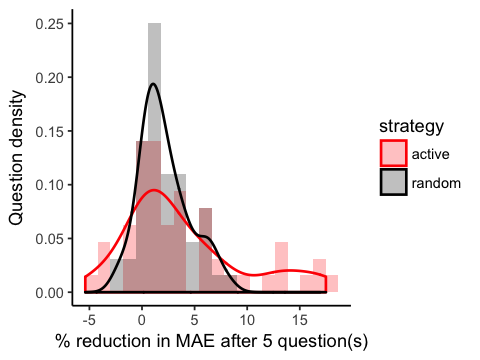}
  \end{subfigure}
  \caption{Reduction in imputation error from pre-survey levels for the initial Facebook survey.}
  \label{fig:facebook-error-reduction}
\end{figure}

\begin{figure}[!htb]
\centering
\includegraphics[width=\textwidth]{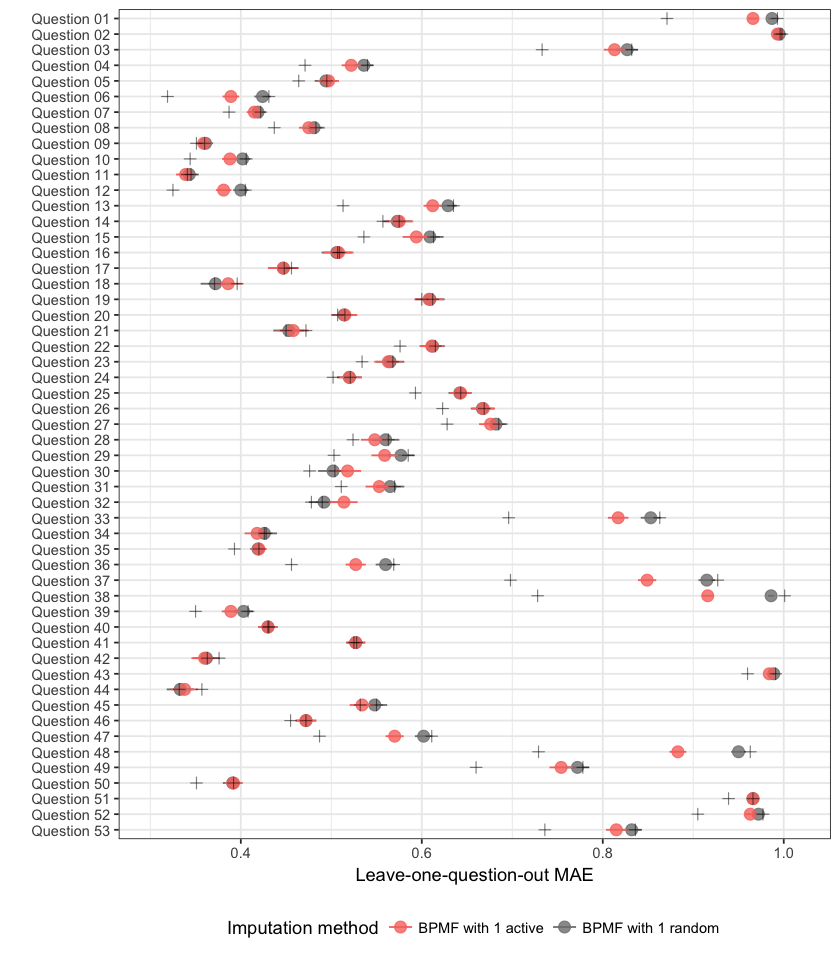}
\caption{Mean absolute prediction error per question after one question chosen actively or randomly, Facebook survey.}
\label{fig:facebook-compare-questions-1}
\end{figure}

\begin{figure}[!htb]
\centering
\includegraphics[width=\textwidth]{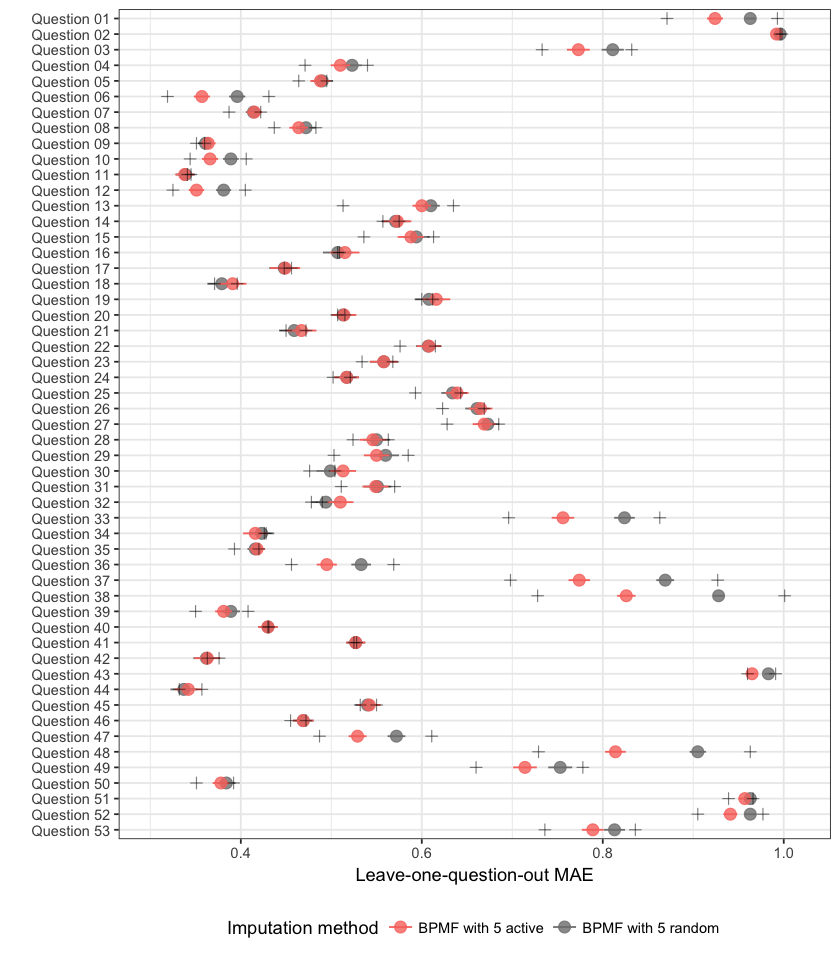}
\caption{Mean absolute prediction error per question after five questions chosen actively or randomly, Facebook survey.}
\label{fig:facebook-compare-questions-5}
\end{figure}

\begin{figure}[!htb]
\centering
\includegraphics[width=\textwidth]{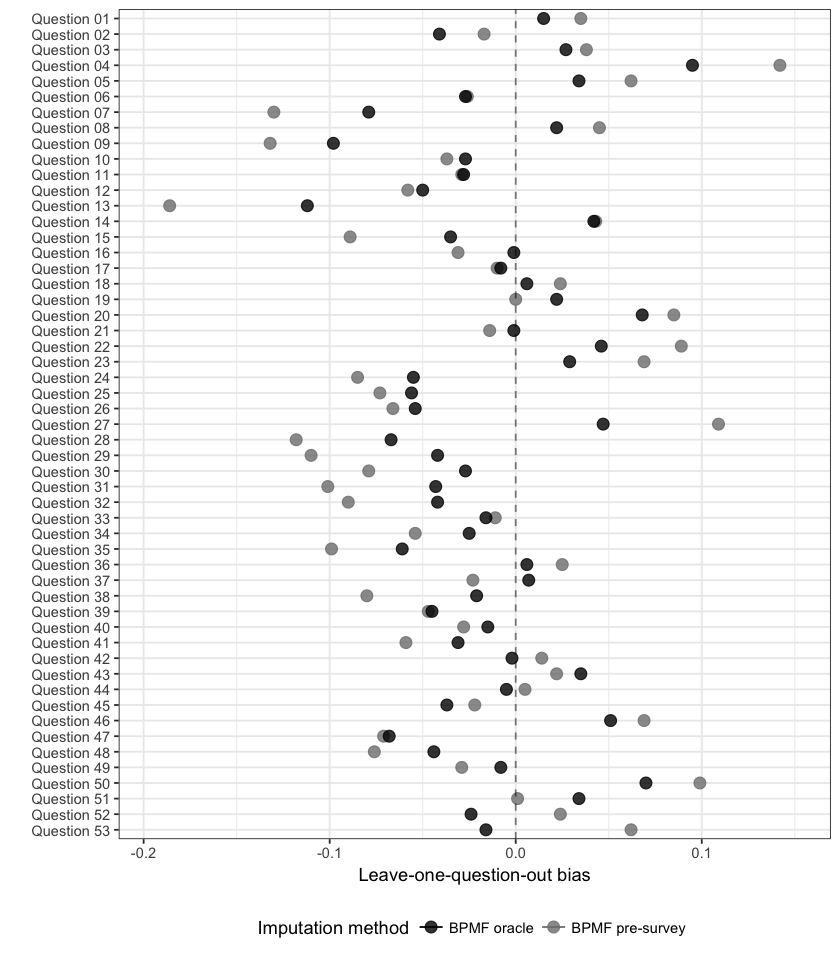}
\caption{Bias per question when imputing responses with no knowledge (``pre-survey'') and with all other responses revealed (``oracle''), Facebook survey.}
\label{fig:facebook-compare-questions-bias}
\end{figure}

\begin{figure}[!htb]
  \centering
  \begin{subfigure}[b]{0.55\textwidth}
    \includegraphics[width=\textwidth]{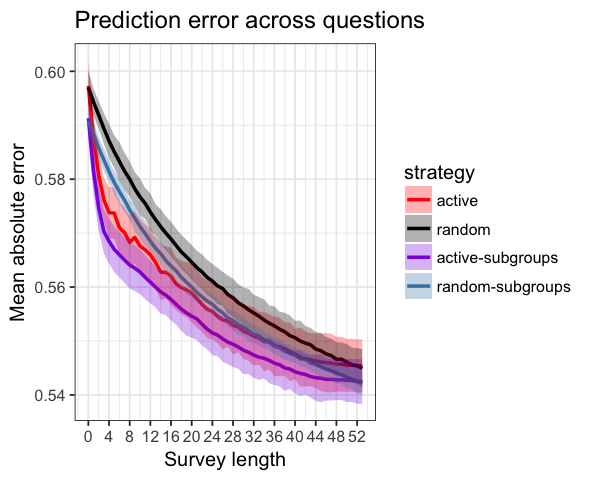}
    \caption{Prediction error on the sparse holdout set, measured across 10 simulations.}
  \end{subfigure}
  \hfill
  \begin{subfigure}[b]{0.40\textwidth}
    \includegraphics[width=\textwidth]{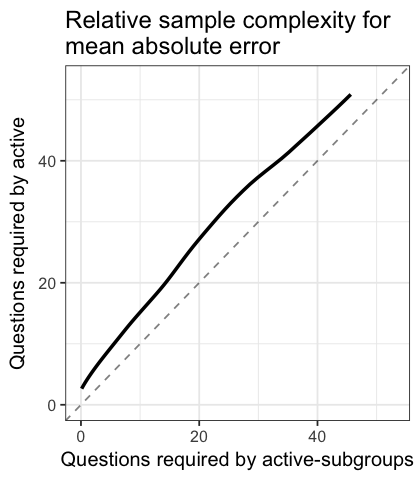}
    \caption{Relative complexity of the active strategy with and without subgroups.}
  \end{subfigure}
  \caption{Summary measures of imputation error for the Facebook survey. The ``-subgroups'' strategies initialize each user prior to the empirical Bayes estimate for the subgroup to which the user belongs. For clarity, we plot $2\sigma$ uncertainty bands across 10 simulations of each strategy.}
  \label{fig:facebook-subgroups-compare-metrics}
\end{figure}

\FloatBarrier

\section{Results for Facebook survey experiment} \label{sec:facebook-exp-results}

\begin{figure}[!htb]
\centering
\includegraphics[width=0.8\textwidth]{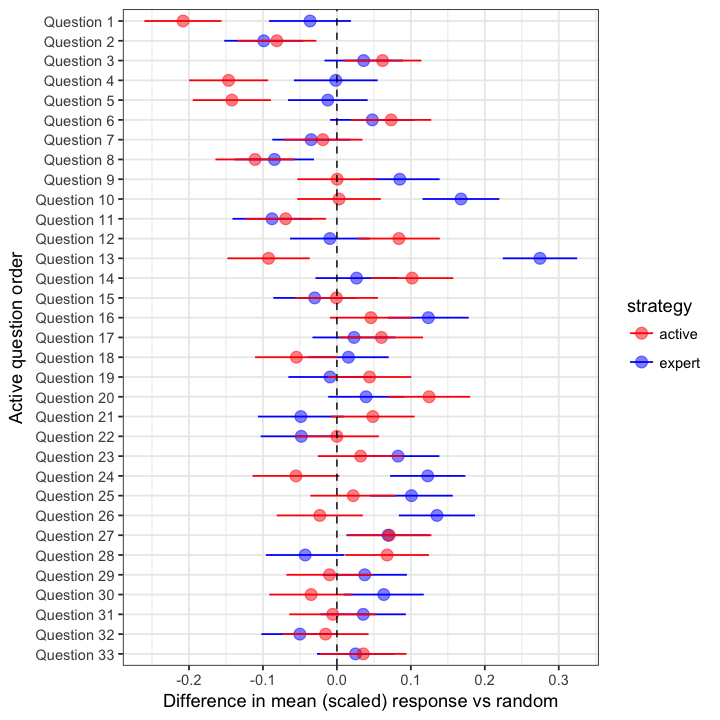}
\caption{Difference in mean response between strategies (conditions), using random order as a baseline, for the Facebook survey experiment. Responses have been rescaled to unit variance on a per-question basis. Hence, as with our estimated order effects, units are standard deviations on the response scale.}
\label{fig:facebook-exp-bias-response}
\end{figure}

\begin{figure}[!htb]
\centering
\includegraphics[width=0.8\textwidth]{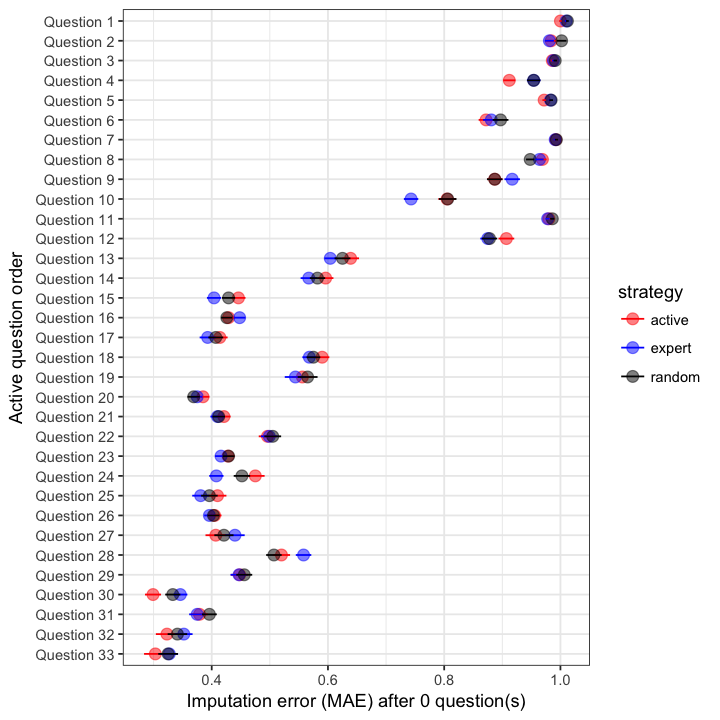}
\caption{Pre-survey imputation error, that is, mean absolute prediction error per question before any responses are available, for the Facebook survey experiment. Predictions are based on the prior mean of user factors and question factors estimated from the initial Facebook survey. Pre-survey imputation error differs across strategies (conditions) due to differences in response distributions induced by order effects and nonresponse.}
\label{fig:facebook-exp-bias-pre-error}
\end{figure}

\begin{figure}[!htb]
  \centering
  \begin{subfigure}{0.7\textwidth}
    \includegraphics[width=\textwidth]{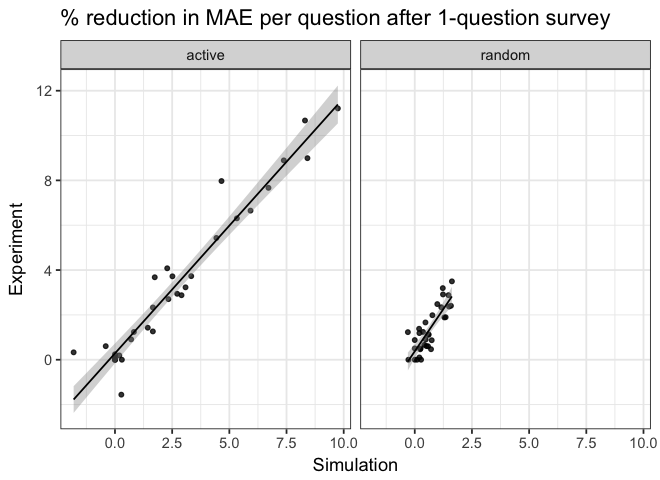}
  \end{subfigure}
  \vskip 5mm
  \begin{subfigure}{0.7\textwidth}
    \includegraphics[width=\textwidth]{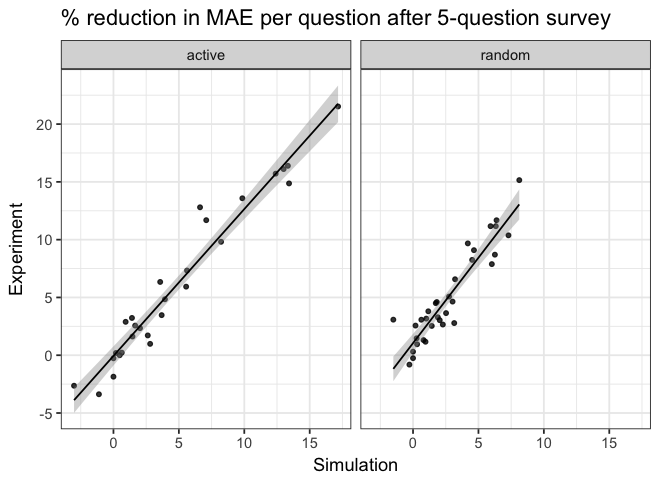}
  \end{subfigure}
  \caption{Comparing simulated and experimental reduction in imputation error on a per-question basis for the Facebook survey. From simulations of the random-order Facebook survey we compute percent reduction in LOOCV mean absolute error from pre-survey levels. From the Facebook survey experiment we compute percent reduction in MAE from pre-survey levels on not-yet-revealed responses. Each point represents one question. We omit questions with undefined experimental imputation error, namely the first question in the active order (top) and the first five questions in the active order (bottom).}
  \label{fig:facebook-exp-versus-simulation}
\end{figure}

\begin{figure}[!htb]
\centering
\includegraphics[width=0.6\textwidth]{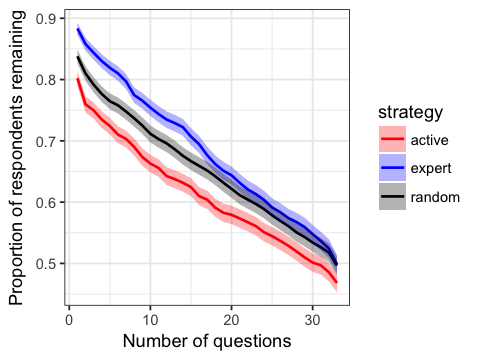}
\caption{Proportion of respondents who remain in the Facebook survey experiment after each question. We show 95\% binomial confidence intervals. After 1 question, the dropout proportion under the active strategy is 3.6\% greater than under the random strategy (95\% confidence interval [1.9\%, 5.2\%]). After 1 question, the dropout proportion under the expert strategy is 4.5\% less than under the random strategy (95\% confidence interval [3.0\%, 6.0\%]).}
\label{fig:facebook-exp-respondents-remaining}
\end{figure}

\FloatBarrier

\section{Results for ordered logit model} \label{sec:ordered-logit-results}


\begin{figure}[!htb]
  \centering
  \begin{subfigure}{0.48\textwidth}
    \includegraphics[width=\textwidth]{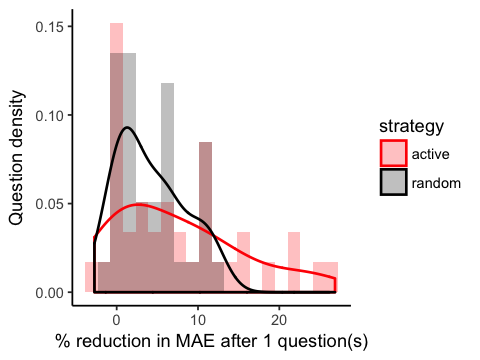}
  \end{subfigure}
  \hfill
  \begin{subfigure}{0.48\textwidth}
    \includegraphics[width=\textwidth]{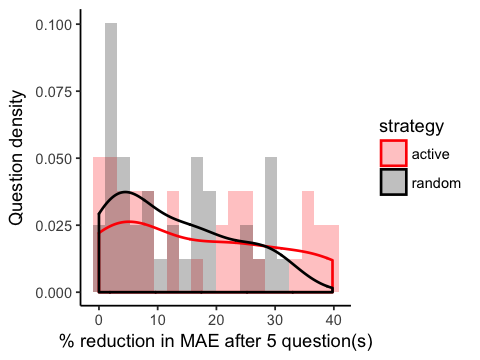}
  \end{subfigure}
  \caption{Reduction in imputation error from pre-survey levels for the 2016 CCES with ordered logit model.}
  \label{fig:cces16-ordlogit-error-reduction}
\end{figure}

\begin{figure}[!htb]
\centering
\includegraphics[width=\textwidth]{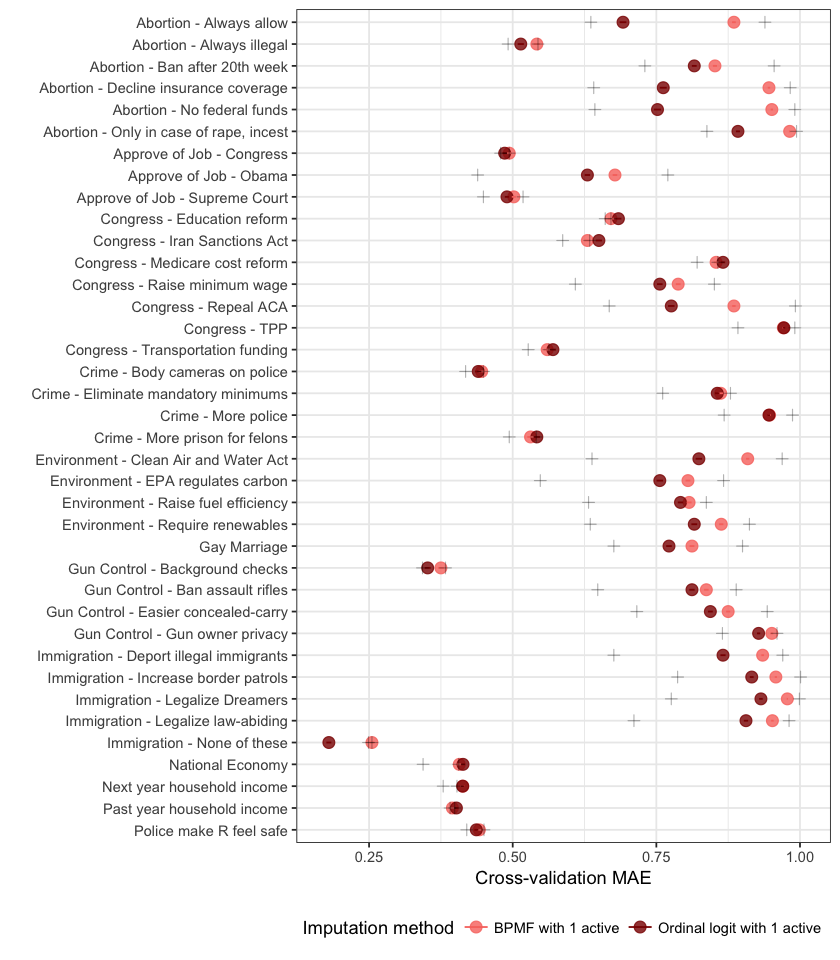}
\caption{Mean absolute prediction error per question after one actively chosen question using PMF or the ordered logit model, CCES 2016. Error is computed by 5-fold cross-validation; PMF and ordered logit use the same folds. Pre-survey and oracle bounds pertain to PMF.}
\label{fig:cces16-ordlogit-vs-bpmf-1}
\end{figure}

\begin{figure}[!htb]
\centering
\includegraphics[width=\textwidth]{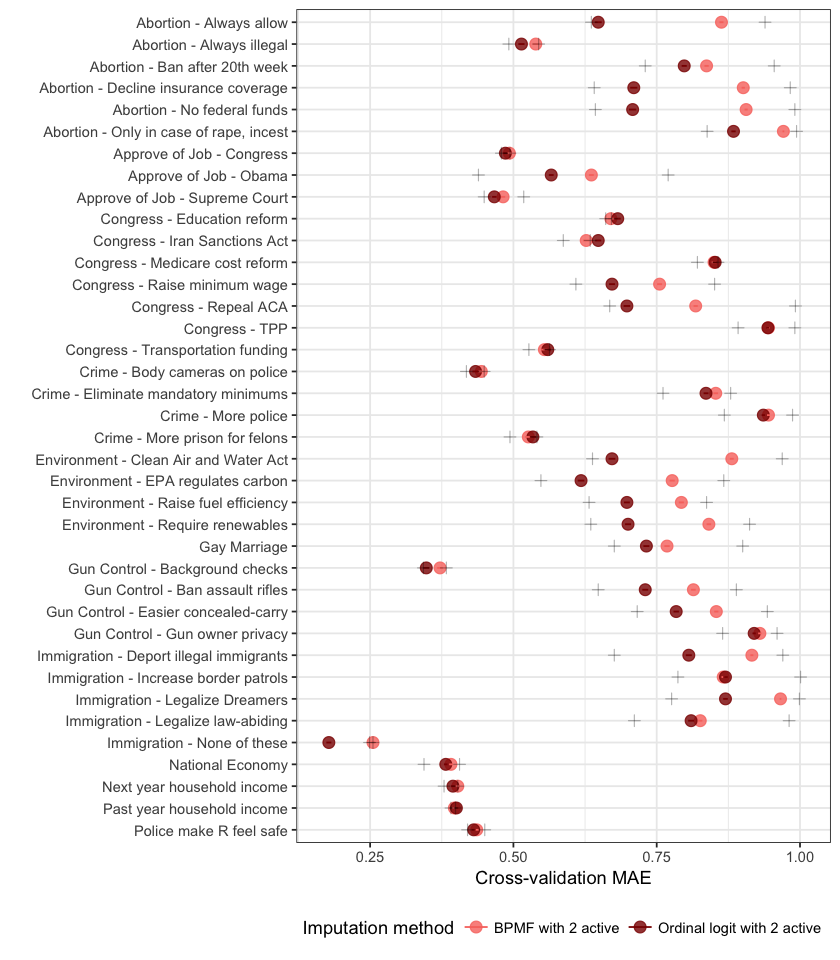}
\caption{Mean absolute prediction error per question after two actively chosen questions using PMF or the ordered logit model, CCES 2016. Error is computed by 5-fold cross-validation; PMF and ordered logit use the same folds. Pre-survey and oracle bounds pertain to PMF.}
\label{fig:cces16-ordlogit-vs-bpmf-2}
\end{figure}


\begin{figure}[!htb]
\centering
\includegraphics[width=\textwidth]{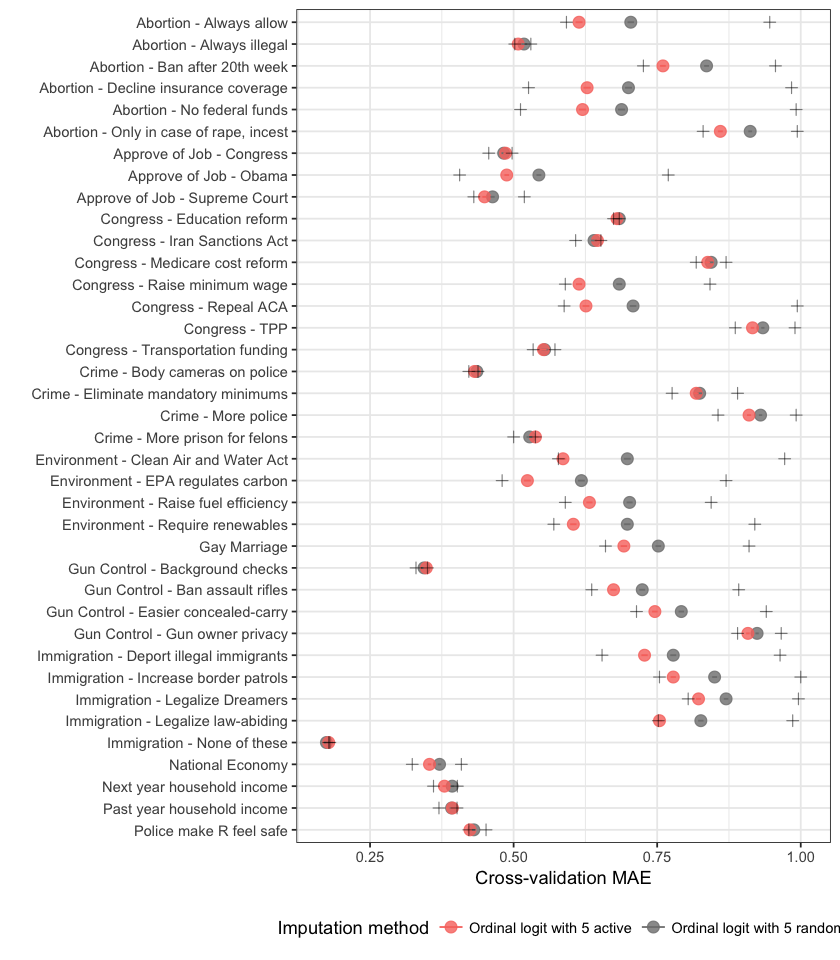}
\caption{Mean absolute prediction error per question after five questions chosen actively or randomly, CCES 2016 with ordered logit model.}
\label{fig:cces16-ordlogit-compare-questions-5}
\end{figure}

\begin{figure}[!htb]
\centering
\includegraphics[width=\textwidth]{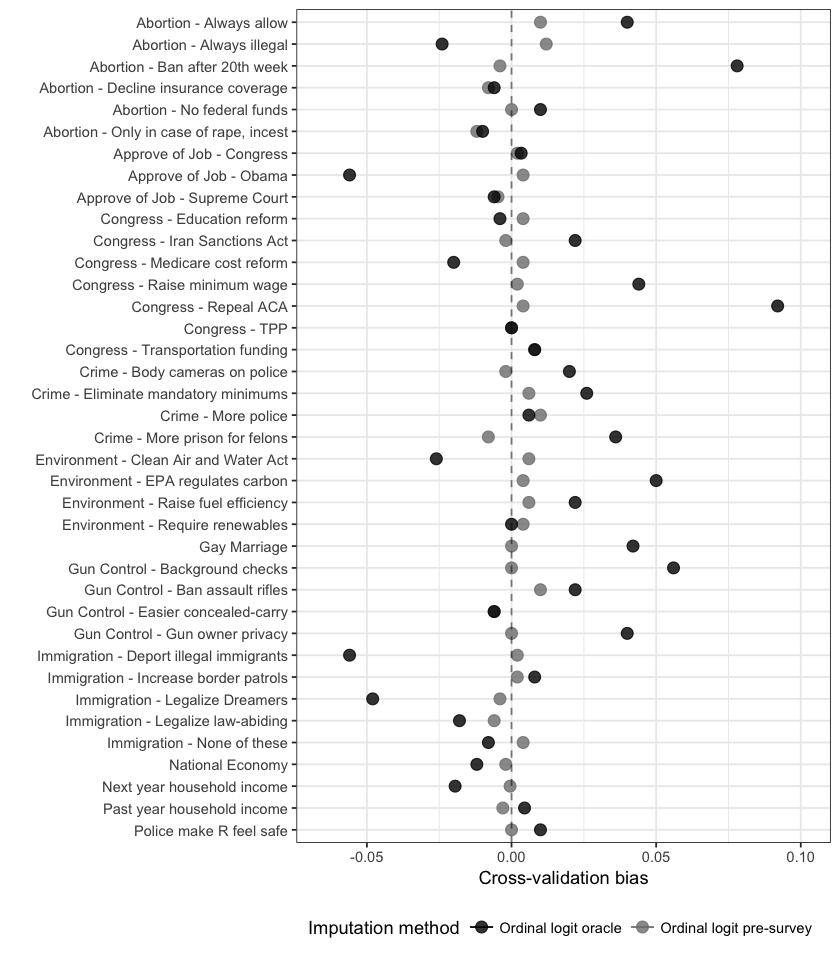}
\caption{Bias per question when imputing responses with no knowledge (``pre-survey'') and with all other responses revealed (``oracle''), CCES 2016 with ordered logit model.}
\label{fig:cces16-ordlogit-compare-questions-bias}
\end{figure}

\begin{figure}[!htb]
\centering
\includegraphics[width=0.8\textwidth]{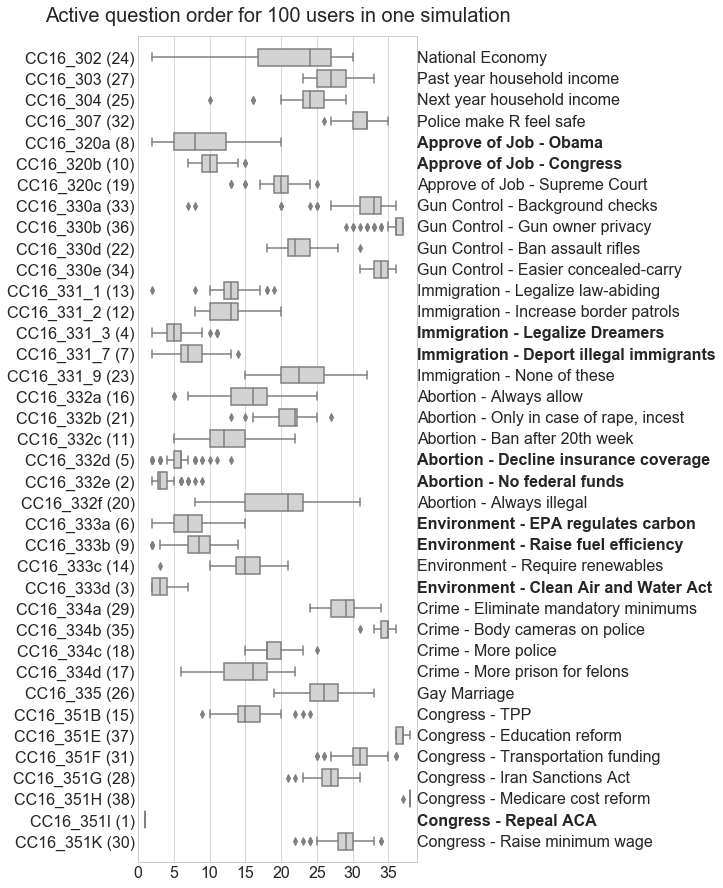}
\caption{Active ordering for 2016 CCES with ordered logit model using a rank-4 matrix decomposition. We show the rank of each question across 100 randomly sampled users in one simulation. \label{fig:cces16-ordlogit-question-rank}}
\end{figure}

\begin{landscape}
\begin{figure}[!htb]
\centering
\includegraphics[scale=0.55]{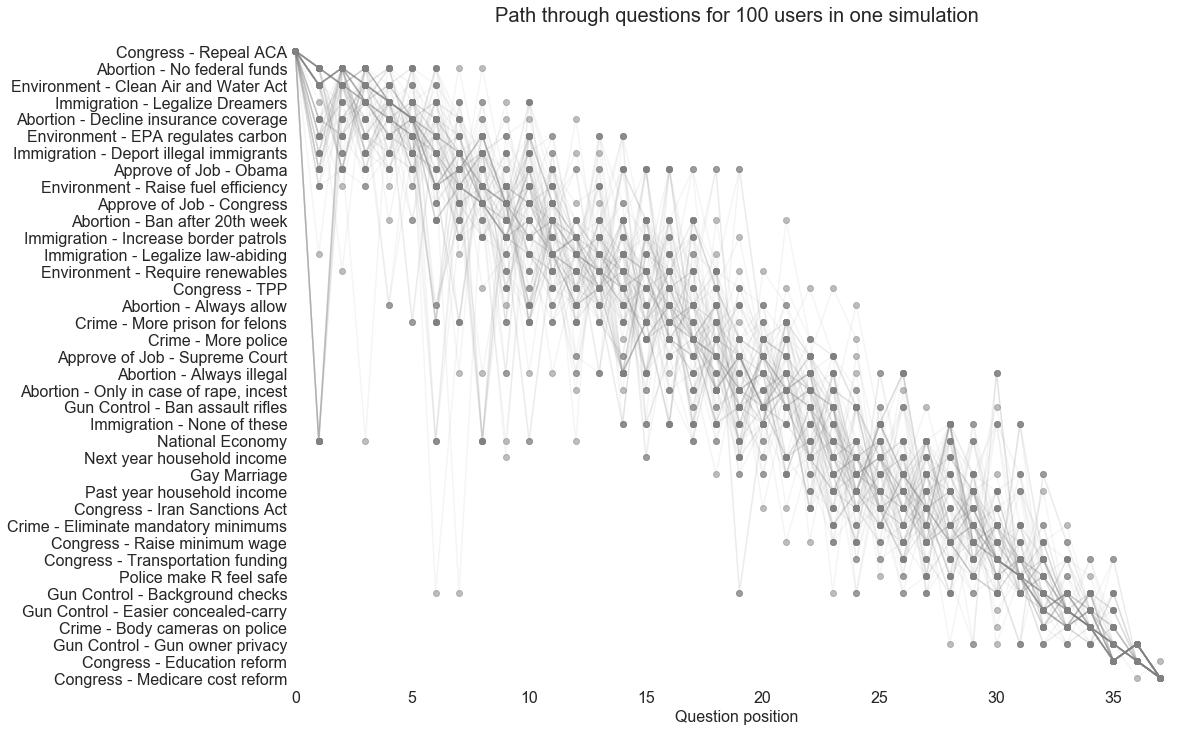}
\caption{Paths through questions on the 2016 CCES, chosen by the active strategy with ordered logit model using a rank-4 matrix decomposition. We show paths for 100 randomly sampled users in one simulation as light gray lines. Darker points indicate questions appearing more frequently at a given position in the survey. \label{fig:cces16-ordlogit-question-paths}}
\end{figure}
\end{landscape}


\end{document}